\documentclass{article}
\usepackage{amsmath,amssymb}
\usepackage{tikz}
\usepackage{graphicx}
\usetikzlibrary{arrows.meta,calc}
\usepackage{microtype}
\usepackage{multirow}
\usepackage{booktabs}
\usepackage{array}
\usepackage{xcolor}
\usepackage{xurl}
\usepackage{algpseudocode}
\usepackage{algorithm}
\usepackage{hyperref}
\usepackage{pgfplots}
\pgfplotsset{compat=1.18}
\usepackage[accepted]{mlsys2025}

\makeatletter
\renewcommand{\mlsys@appearing}{\textit{Preprint. November 2026.}}
\makeatother

\newcommand{\R}{\mathbb{R}}

\makeatletter
\newcommand{\tabdashline}[1]{%
  \multicolumn{#1}{@{}l@{}}{%
    \xleaders\hbox to 5pt{\hss\raise\dimexpr\ht\@arstrutbox-\arrayrulewidth\relax
      \hbox{\vrule width 2pt height \arrayrulewidth}\hss}\hfill\kern0pt}%
  \\[\dimexpr-\ht\@arstrutbox-\dp\@arstrutbox+\arrayrulewidth\relax]}
\newcommand{\tabdashvline}[1][0pt]{%
  \kern2pt%
  \lower\dp\@arstrutbox\vbox to \dimexpr\ht\@arstrutbox+\dp\@arstrutbox+#1\relax{%
    \offinterlineskip
    \xleaders\vbox to 4pt{\vfil\hrule width \arrayrulewidth height 2pt\vfil}\vfill
  }%
  \kern2pt%
}
\makeatother

\hypersetup{
    pdftitle={Breaking Routing And Inter-device Dependencies in Mixture-of-Experts},
    pdfauthor={John Zarka, Danyal Ahmed, Eugene Belilovsky, Edouard Oyallon},
    pdfsubject={Preprint},
    pdfkeywords={Mixture-of-Experts, Expert Parallelism, Early Routing, Ladder Residual, Communication-Computation Overlap, LLM Inference}
}

\mlsystitlerunning{Breaking Routing And Inter-device Dependencies in Mixture-of-Experts}

\begin{document}

\twocolumn[
\mlsystitle{Breaking Routing And Inter-device Dependencies in Mixture-of-Experts}

\mlsyssetsymbol{equal}{*}

\begin{mlsysauthorlist}
\mlsysauthor{John Zarka}{equal,su}
\mlsysauthor{Danyal Ahmed}{equal,su}
\mlsysauthor{Eugene Belilovsky}{ca,mila}
\mlsysauthor{Edouard Oyallon}{su,cnrs}
\end{mlsysauthorlist}

\mlsysaffiliation{su}{ISIR, Sorbonne Université, Paris, France}
\mlsysaffiliation{ca}{Department of Computer Science and Software Engineering, Concordia University, Montréal, Canada}
\mlsysaffiliation{mila}{Mila -- Quebec AI Institute, Montréal, Canada}
\mlsysaffiliation{cnrs}{CNRS, Paris, France}

\mlsyscorrespondingauthor{John Zarka}{john.zarka@sorbonne-universite.fr}

\vskip 0.3in

\begin{abstract}
As Mixture-of-Experts (MoE) models scale, Expert Parallelism (EP) becomes increasingly important for distributing experts across devices. However, the resulting token dispatch and combine operations can become a major bottleneck during inference. Overlapping these communication phases with computation is constrained by the sequential dependencies of standard MoE architectures. We introduce BRAID-MoE, an architecture that pairs laddered residual connections with early routing to relax these dependencies. Laddered residuals enable each expert combine to be deferred to the following block, while early routing computes the next routing decision from the pre-combine residual, with experts still receiving the updated post-combine hidden state. This creates an extended communication window in which the preceding combine and next dispatch can execute back-to-back on a dedicated communication stream, overlapping attention computation. On Qwen3-shaped MoEs with 0.4B-A0.1B and 2.5B-A0.5B parameters, BRAID closely matches sequential training at matched compute, with validation-loss increases of only 1\% and 0.5\%, respectively. We further demonstrate that existing MoE checkpoints can be adapted to BRAID with limited additional training, largely recovering their original accuracy. On a converted Qwen3-30B-A3B model, BRAID achieves speedups of up to $1.29\times$ in prefill and $1.37\times$ in decode, and at least $1.12\times$ across diverse workloads and EP configurations on one or two four-H100 nodes. It outperforms competing overlap strategies in all tested settings, with the strongest gains when communication is a major latency bottleneck and its duration approaches the available computation window. These results demonstrate that early routing enables more effective communication--computation overlap in MoE inference at a modest cost in model quality.
\end{abstract}
]

\printAffiliationsAndNotice{\mlsysEqualContribution}

\section{Introduction}
Sparse MoE models increase parameter capacity without increasing per-token arithmetic proportionally: each token activates only a small subset of a much larger expert bank~\citep{shazeer2017outrageously,lepikhin2021gshard,fedus2022switch}. This design has become central to frontier open-weight models. Qwen3, for example, includes MoE variants up to 235B total parameters with 22B activated per token~\citep{qwen2025qwen3}; DeepSeek-V3 uses a large routed expert bank with sparse activation~\citep{deepseekai2025deepseekv3technicalreport}; and Kimi K3 pushes sparsity further, activating 16 of 896 routed experts inside a 2.8T-parameter model~\citep{kimi2026k3}. As the expert bank grows faster than the active computation, EP becomes essential for scaling MoEs: experts are sharded across devices, and tokens move to the devices hosting their selected experts. 

Under EP, each MoE layer dispatches tokens to remote experts and combines the returned outputs. These communication phases can account for a substantial fraction of execution time. For instance, COMET~\citep{zhang2025comet} reports that inter-device communication accounts for nearly half of total execution time on average for several popular MoEs. Moreover, sequential dependencies between communication and computation limit opportunities for overlap. To mitigate this overhead, specialized libraries such as DeepEP~\citep{deepseek2025deepep} provide high-throughput, low-latency dispatch and combine kernels, while COMET~\citep{zhang2025comet} optimizes communication--computation overlap and Lina~\citep{li2023lina} addresses communication contention and imbalanced all-to-all traffic. A common approach combines low-level kernel optimizations with a dual-stream design that enables concurrent communication and computation, balancing SM resource usage and minimizing unnecessary synchronization to reduce overall execution time.

While communication--computation overlap has been extensively explored in MoE training, inference presents a different set of challenges. Training offers additional scheduling flexibility~\citep{guo2026sonicmoe}, exploiting independent backward computations~\citep{jiang2024lancet} and forward--backward pipelining across microbatches~\citep{deepseekai2025deepseekv3technicalreport} to hide communication latency. Inference, however, is restricted to the forward pass, where dependencies between expert dispatch, computation, and combination leave fewer opportunities for overlap. Exploiting parallelism therefore often requires partitioning tokens into microbatches and pipelining their execution~\citep{deepseekai2025deepseekv3technicalreport,vllm_dbo}. Yet, the effectiveness of such approaches depends heavily on the available token volume. During prefill, long input sequences provide sufficient tokens for chunking, whereas autoregressive decoding generates only one token per request at each step. Consequently, hiding EP communication during decoding typically requires batching multiple concurrent requests, which can be challenging under low concurrency, memory constraints, or strict latency requirements.

Architectural modifications offer another way to improve MoE execution efficiency. Although such changes require ensuring that model quality is preserved, they can expose new opportunities for system-level optimization. Our work builds on two complementary ideas: ladder architectures and early routing. Ladder architectures~\citep{zhang2025ladder,dukler2026farskip} introduce minor modifications to the Transformer architecture by duplicating the residual stream and shifting attention and FFN/MoE computations relative to one another, enabling communication--computation overlap with minimal impact on model quality. This approach is not specific to MoEs and was initially explored for overlapping tensor-parallel communication with computation in dense Transformers. Moreover, \citet{zhang2025ladder} demonstrated that standard Transformer models can be converted to ladder architectures with limited additional training. Separately, early routing~\citep{hwang2024pregated,cai2025shortcut} makes expert assignments available in earlier layers, allowing token dispatch to begin before the corresponding MoE layer and thereby reducing communication latency, at the potential cost of degraded model convergence. 

Building on these complementary approaches, we introduce BRAID-MoE (Breaking Routing And Inter-device Dependencies in Mixture-of-Experts), which combines laddered residual connections and early routing to expose additional communication--computation overlap while mitigating their respective limitations. BRAID decouples routing from the preceding combine operation while preserving fresh activations for expert computation, allowing consecutive combine and dispatch operations across layer boundaries to be executed back-to-back without an intervening routing step. We exploit this architecture through a two-stream execution schedule using DeepEP asynchronous communication kernels, enabling both communication phases to overlap with attention. Furthermore, by scheduling expert computation after these communication operations, BRAID creates opportunities to overlap expert execution with subsequent attention-side communication, such as tensor-parallel collectives.

\paragraph{Contributions.}
Our contributions are as follows: \\
\textbf{(i)} We introduce BRAID-MoE, an architecture that pairs laddered residual connections with early routing. The router reads the pre-combine residual, while dispatch and expert computation use fresh post-combine activations. This removes routing from the serial combine--dispatch path, opening the possibility of future fusion of both communication phases into a single kernel.
\textbf{(ii)} We implement an event-driven two-stream execution schedule and a custom DeepEP combine epilogue that fuses residual addition and RMSNorm. The schedule executes combine and dispatch as back-to-back communication kernels, overlapping both phases with routing and attention. 
\textbf{(iii)} We show that BRAID can be pretrained from scratch at scales up to 2.5B-A0.5B parameters with less than a 1\% increase in validation cross-entropy relative to matched sequential MoEs.
\textbf{(iv)} We demonstrate that pretrained sequential MoE checkpoints can be converted to BRAID through limited continued training, including post-training recovery on an instruction-tuned OLMoE-1B-7B checkpoint.
\textbf{(v)} We achieve speedups of up to $1.29\times$ in prefill and $1.37\times$ in decode over sequential execution on a converted Qwen3-30B-A3B checkpoint, using one or two four-H100 nodes, and outperform competing communication--computation overlap schedules in all tested settings.

\paragraph{Structure of the paper.} Section~\ref{sec:method} introduces the Sequential, Parallel, Ladder, and Early-Routing MoE formulations. Section~\ref{sec:braid} presents the BRAID architecture and its multi-stream execution schedule. Section~\ref{sec:quality} evaluates pretraining, checkpoint adaptation, and post-training recovery, while Section~\ref{sec:perf} reports prefill and decode latency and communication overlap. Our source code, based on TorchTitan\footnote{\url{https://github.com/pytorch/torchtitan}}~\citep{liang2025torchtitan} and DeepEP\footnote{\url{https://github.com/deepseek-ai/DeepEP}}~\citep{deepseek2025deepep}, can be found at \url{https://github.com/j-zarka/braid-moe/}.

\section{Related Work}
\label{sec:related}

\paragraph{Reducing communication volume.}

One approach to improving EP communication efficiency is to reduce the amount of data exchanged between devices. LatentMoE~\citep{elango2026latentmoe} introduces shared down- and up-projections around the routed expert path, allowing token dispatch, expert computation, and output combination to operate in a lower-dimensional latent space~\citep{janson2026stabilizing,janson2026learned}. This reduces both communication volume and expert memory traffic, enabling greater expert capacity at comparable execution cost. More recently, Kimi K3 adopted Stable LatentMoE, reducing the routed representation from 7168 to 3584 dimensions while activating 16 out of 896 routed experts~\citep{kimi2026k3}. Multi-Head LatentMoE further explores architectural changes to improve communication efficiency through head parallelism~\citep{cui2026multihead}. We also note that local SGD methods could be combined with EP~\citep{belilovsky2026inner,cagnasso2026unifyinglocalcommunicationslocal,cagnasso2026spicesparseparametergossip}, although they are not specifically designed for MoEs. These approaches aim to reduce communication costs, whereas our work focuses on hiding communication latency through overlap.

\paragraph{EP optimizations and multi-stream execution.}
Another line of work improves EP execution without modifying the model architecture or changing its expressivity. DeepEP~\citep{deepseek2025deepep} provides specialized high-throughput and low-latency kernels for token dispatch and combine, supporting asynchronous communication. Tutel~\citep{hwang2023tutel} introduces adaptive parallelism and communication--computation pipelining to accommodate varying expert workloads, while Lina~\citep{li2023lina} addresses communication contention during training and imbalanced all-to-all traffic during inference. More generally, multi-stream implementations schedule communication and computation on separate GPU streams to enable concurrent execution, although overlapping kernels may compete for GPU resources. Such overlap is also limited by the dependencies of the underlying model: communication can only overlap with operations whose inputs are already available.

\paragraph{Communication--computation overlap and scheduling.}
Several systems improve overlap by decomposing and rescheduling operations within the existing computation graph. COMET~\citep{zhang2025comet} uses fine-grained dependency analysis and workload scheduling to overlap MoE communication with expert computation. FLUX~\citep{chang2024flux} fuses communication with neighboring GPU computation, primarily targeting tensor-parallel collectives. At the graph level, Lancet~\citep{jiang2024lancet} exploits additional scheduling opportunities during training, notably by overlapping communication with backward-pass computations. We note that additional overlap can also be enabled by modifying the backward pass~\citep{rivaud2025petra,nabli2025acco,kopyshevskiy2026}, although this approach is specific to training. Microbatching provides another avenue for overlap during inference: DeepSeek-V3 schedules attention and MoE execution from different microbatches concurrently~\citep{deepseekai2025deepseekv3technicalreport}, while vLLM's Dual Batch Overlap interleaves communication and computation across successive microbatches~\citep{vllm_dbo}. While effective, these techniques are constrained by the sequential dependencies of the underlying MoE architecture, which limit opportunities to expose independent computation and overlap communication.

\paragraph{Architectural parallelism through residual connections.}
Beyond execution-level optimizations, architectural modifications can directly relax dependencies between Transformer sublayers. Architectures with parallel attention and FFN branches, notably used in PaLM~\citep{chowdhery2023palm}, allow both branches to operate on the same residual state, exposing additional concurrency. Ladder Residual~\citep{zhang2025ladder} extends this idea by having each module operate on an earlier residual state, allowing its computation to overlap with communication from the preceding module. Although applicable to general residual architectures, Ladder Residual was originally developed to overlap tensor-parallel communication with computation in dense Transformers. The authors also demonstrate that existing pretrained Transformers can be partially converted to this architecture with limited additional training. Extending the principle to MoEs, FarSkip-Collective~\citep{dukler2026farskip} uses partially updated residual states to allow the next attention layer to begin before the preceding MoE combine completes, overlapping both combine and dispatch with attention computation. However, routing still depends on the preceding combine, preserving a sequential combine--routing--dispatch dependency and making the effectiveness of overlap sensitive to the partitioning of attention computation. These works demonstrate how architectural modifications can expose additional parallelism for system-level optimizations to exploit.

\paragraph{Anticipating routing decisions.}
A complementary approach is to anticipate routing decisions by computing expert assignments from earlier representations. Pre-gated MoE~\citep{hwang2024pregated} evaluates routing decisions one MoE block ahead, allowing selected expert weights to be prefetched from CPU to GPU while the preceding MoE block executes. Although motivated by expert offloading rather than overlapping EP communication, it demonstrates the potential of computing routing decisions earlier. RMoE~\citep{qiu2025rmoe} explores a related but distinct approach through a recurrent router, leveraging cross-layer information to improve expert selection. More closely related to our setting, ScMoE~\citep{cai2025shortcut} introduces cross-layer shortcuts to overlap expert communication with attention and FFN computation. In its main configuration, dense FFN and MoE blocks are interleaved, with the routed experts consuming representations from the preceding block. This creates a larger window for communication--computation overlap, but both routing and routed-expert computation rely on earlier activations. In contrast, BRAID decouples routing from the preceding combine operation: expert assignments are computed from the pre-combine residual, while dispatch operates on fresh post-combine activations. Combined with laddered residual connections, this allows combine and dispatch to execute back-to-back, enabling more efficient overlap with attention and opening the possibility of fusing both communication phases into a single kernel. It also enables expert computation to overlap with subsequent attention-side communication.

\section{Existing Methods}
\label{sec:method}
We now introduce the architectural variants considered in this work, highlighting how each modifies the dependencies between attention, routing, and expert computation to expose communication--computation overlap. Figure~\ref{fig:architectures} illustrates the different configurations discussed below.

\subsection{Sequential MoEs}
\label{sec:seq-moe}

We index Transformer blocks by $\ell$ and denote the number of tokens by $T$. Following \citet{guo2026sonicmoe}, let $d$ denote the hidden dimension, $E$ the number of routed experts, and $K$ the number of experts selected per token. For readability, we adopt the pre-normalization convention and omit explicit normalization operators, implicitly incorporating them into the corresponding attention and MoE operations. Given the input $h_\ell\in\R^{T\times d}$ to block $\ell$, a standard MoE Transformer~\citep{lepikhin2021gshard,fedus2022switch} performs:
\begin{equation}
\begin{aligned}
z_\ell
    &= h_\ell + \operatorname{Attn}_\ell(h_\ell),\\
(S_\ell,\pi_\ell)
    &= \operatorname{Route}_\ell(z_\ell),\\
h_{\ell+1}
    &= z_\ell + \operatorname{MoE}_\ell(z_\ell;S_\ell,\pi_\ell)\,,
\end{aligned}
\label{eq:standard-moe}
\end{equation}
where, for $i\in[T]$,
\begin{equation*}
\operatorname{MoE}_\ell(z_\ell;S_\ell,\pi_\ell)_i \triangleq\sum_{e=1}^{E}       \pi_{\ell,i,e}\,S_{\ell,i,e}\,\operatorname{Expert}_{\ell,e}
       (z_{\ell,i}).
\end{equation*}
Here, $z_\ell\in\R^{T\times d}$ is the post-attention representation, $\operatorname{Attn}_\ell$ denotes the attention submodule of block $\ell$, and $\operatorname{Expert}_{\ell,e}$ the FFN corresponding to expert $e$ in layer $\ell$. The matrix $S_\ell\in\R^{T\times E}$ contains the routing weights, while $\pi_\ell\in\{0,1\}^{T\times E}$ encodes the top-$K$ expert assignments, with $\sum_e\pi_{\ell,i,e}=K$.

\paragraph{Expert Parallelism.} Under EP, experts are distributed across devices. Let $\mathcal{J}_r\subseteq[E]$ denote the experts hosted on rank $r$. The gather operation selects the tokens assigned to each expert. Under EP, this becomes a distributed \emph{dispatch} operation $\operatorname{Dispatch}_{\ell}(\cdot,\pi_\ell)$, which transfers tokens to the devices hosting their selected experts. Thus, during expert computation, each rank holds a local set of experts
\begin{equation*}
\operatorname{Expert}^{\mathrm{loc}}_{\ell,r}
    = \{\operatorname{Expert}_{\ell,e}:e\in\mathcal{J}_r\}.
\end{equation*}
A subsequent \emph{combine} returns these experts' outputs to their source ranks and performs the routing-weighted aggregation to obtain the MoE output. Until combine, these outputs remain sharded across the expert-hosting ranks. Let $\mu_\ell$ denote the routing weights and dispatch metadata retained for combining the expert outputs of layer $\ell$; we write $\operatorname{Combine}_\ell(\cdot;\mu_\ell)$ for the corresponding collective. For clarity, we abbreviate $\operatorname{Route}$, $\operatorname{Attn}$, $\operatorname{Expert}$, $\operatorname{Combine}$, and $\operatorname{Dispatch}$ as $R$, $A$, $\mathcal{E}$, $C$, and $D$, respectively, throughout the diagrams.

\subsection{Parallel MoEs}
\label{sec:para-moe}

A simple way to expose additional parallelism is to evaluate attention and MoE from the same residual state, rather than sequentially as in Eq.~\eqref{eq:standard-moe}. This parallel formulation, following the parallel Transformer design~\citep{chowdhery2023palm}, removes the dependency of expert computation on the current attention output. A parallel MoE block is then written as
\begin{equation*}
\begin{aligned}
(S_\ell,\pi_\ell)
    &= \operatorname{Route}_\ell(h_\ell),\\
h_{\ell+1}
    &= h_\ell +\operatorname{Attn}_\ell(h_\ell)+\operatorname{MoE}_\ell(h_\ell;S_\ell,\pi_\ell)\,.
\end{aligned}
\end{equation*}

Since both branches consume $h_\ell$, attention can overlap with expert dispatch once routing completes. However, expert computation is scheduled after attention, as overlapping the two can cause GPU resource contention (\hyperref[par:contention]{Appendix~\ref*{app:tuning} (SM allocation and resource contention)}). The next block still depends on the completion of both branches, including the MoE combine operation. Thus, while parallel MoEs expose intra-block overlap, communication remains on the critical path between successive blocks.

\subsection{Ladder MoEs}
\label{sec:ladd-moe}

Ladder Residual~\citep{zhang2025ladder} further relaxes the dependencies between consecutive Transformer sublayers by introducing two interleaved residual streams. Instead of consuming the most recently updated residual, each sublayer operates on an earlier representation, while its output is accumulated into the current residual stream. In the MoE setting, a formulation closely related to that of \citet{dukler2026farskip} is
\begin{equation*}
\begin{aligned}
(S_\ell,\pi_\ell)
    &= \operatorname{Route}_\ell(y_\ell),\\
h_{\ell+1}
    &= y_\ell + \operatorname{Attn}_\ell(h_\ell),\\
y_{\ell+1}
    &= h_{\ell+1}
       + \operatorname{MoE}_\ell(y_\ell;S_\ell,\pi_\ell).
\end{aligned}
\end{equation*}
Here, $y_\ell$ denotes the residual after incorporating the preceding MoE output. Unlike parallel MoEs, the two branches now operate on different residual states. In particular, attention consumes $h_\ell$, which is available before the preceding MoE combine completes, while routing and expert computation consume the updated residual $y_\ell$. This allows attention in layer $\ell$ to overlap with the preceding combine operation $C_{\ell-1}$.

However, routing requires $y_\ell$, which is available only after the preceding combine completes. Dispatch must therefore wait for routing, preserving a sequential combine--route--dispatch dependency that limits further communication overlap.

\subsection{Early-Routing MoEs}
\label{sec:early-ladder}

Separately, \citet{hwang2024pregated} propose computing routing decisions from the earlier representation $z_{\ell-1}$, while preserving the fresh representation $z_\ell$ for expert computation. A $p$-Early-Routing MoE computes routing decisions from a residual representation $p$ layers earlier, while otherwise retaining the standard MoE formulation in Eq.~\eqref{eq:standard-moe}. Specifically, the routing equation becomes
\begin{equation*}
(S_\ell,\pi_\ell)
    = \operatorname{Route}_\ell(z_{\ell-p}).
\end{equation*}
For $p=1$, since $z_{\ell-1}$ is available before the previous MoE combine completes, expert assignments can be computed in advance. This removes routing from the combine-to-dispatch critical path. However, dispatch still requires the updated post-attention representation $z_\ell$, preserving a combine--attention--dispatch dependency that prevents back-to-back communication.

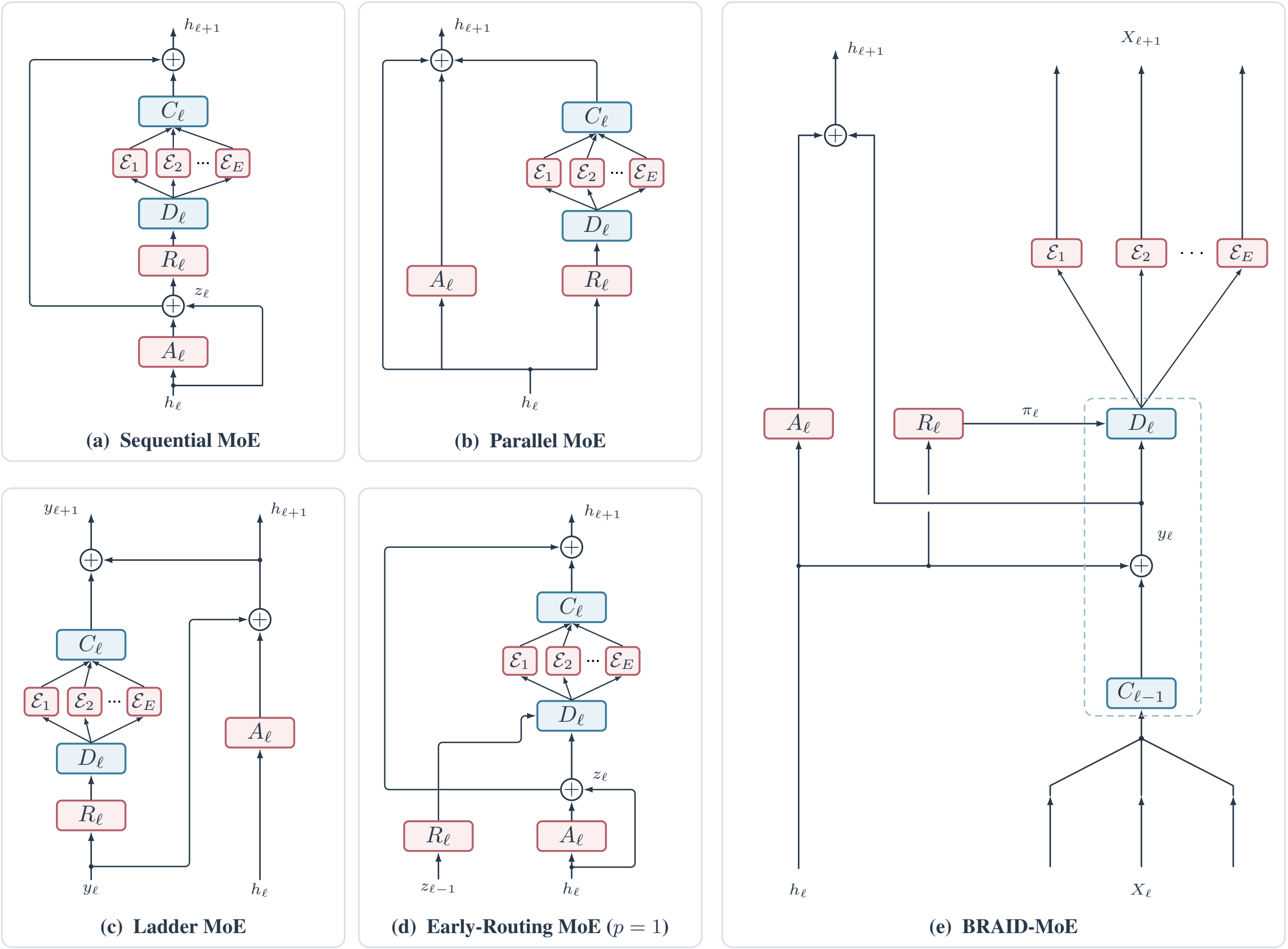
\begin{figure*}[t]
    \centering
    \resizebox{\textwidth}{!}{%
\begingroup
\definecolor{moeink}{HTML}{283849}
\definecolor{moemuted}{HTML}{66798B}
\definecolor{moeborder}{HTML}{D8E0E7}
\definecolor{moered}{HTML}{B65F69}
\definecolor{moeredfill}{HTML}{FBEFF0}
\definecolor{moeblue}{HTML}{397C9D}
\definecolor{moebluefill}{HTML}{E8F2F7}
\begin{tikzpicture}[
    x=1cm,y=1cm, font=\footnotesize,
    >=Latex,
    flow/.style={draw=moeink,line width=.66pt,-{Latex[length=1.5mm,width=1mm]},line cap=round,line join=round},
    pathline/.style={draw=moeink,line width=.66pt,line cap=round,line join=round},
    bypass/.style={draw=moeink,line width=.61pt,-{Latex[length=1.45mm,width=.95mm]},rounded corners=1.8pt},
    comp/.style={rectangle,rounded corners=2.0pt,draw=moered,fill=moeredfill,
        line width=.84pt,minimum width=1.06cm,minimum height=.46cm,inner sep=1pt,
        font=\normalsize,text=moeink},
    comm/.style={rectangle,rounded corners=2.0pt,draw=moeblue,fill=moebluefill,
        line width=.84pt,minimum width=1.06cm,minimum height=.46cm,inner sep=1pt,
        font=\normalsize,text=moeink},
    expert/.style={comp,minimum width=.52cm,minimum height=.43cm,font=\small},
    plus/.style={circle,draw=moeink,fill=white,line width=.72pt,
        inner sep=0pt,minimum size=.33cm,font=\small,text=moeink},
    state/.style={font=\scriptsize,text=moeink,inner sep=1pt},
    commregion/.style={rounded corners=4pt,draw=moeblue!50,densely dashed,line width=.66pt},
    frame/.style={rounded corners=4.5pt,draw=moeborder,fill=white,line width=.7pt},
    fan/.style={draw=moeink,line width=.56pt,-{Latex[length=1.20mm,width=.82mm]}},
]
\def\moepanel#1#2{%
    \draw[frame] (-2.65,-.70) rectangle (2.65,6.40);
    \node[font=\small\bfseries,text=moeink] at (0,-.40) {(#1)\enspace #2};
}
\def\moedots#1#2{%
  \foreach \dotfraction in {0.30,0.50,0.70} {%
    \fill[black] ($(#1.east)!\dotfraction!(#2.west)$) circle (.017cm);%
  }%
}
\def\moeexperts#1#2#3{%
  \node[expert] (#1Eone) at ($(#2,#3)+(-.67,0)$) {$\mathcal{E}_1$};
  \node[expert] (#1Etwo) at (#2,#3) {$\mathcal{E}_2$};
  \node[expert] (#1Ek) at ($(#2,#3)+(.92,0)$) {$\mathcal{E}_E$};
  \moedots{#1Etwo}{#1Ek}%
}
\def\moefanout#1#2{%
    \draw[fan] (#1.north) -- (#2Eone.south);
    \draw[fan] (#1.north) -- (#2Etwo.south);
    \draw[fan] (#1.north) -- (#2Ek.south);
}
\def\moefanin#1#2{%
    \draw[fan] (#1Eone.north) -- (#2.south);
    \draw[fan] (#1Etwo.north) -- (#2.south);
    \draw[fan] (#1Ek.north) -- (#2.south);
}

\begin{scope}[shift={(2.65,7.52)}]
  \moepanel{a}{Sequential MoE}
  \node[comp] (aA) at (0,.99) {$A_\ell$};
  \node[plus] (aAttSum) at (0,1.70) {$+$};
  \node[comp] (aR) at (0,2.40) {$R_\ell$};
  \node[comm] (aD) at (0,3.13) {$D_\ell$};
  \moeexperts{a}{0}{3.91}
  \node[comm] (aC) at (0,4.71) {$C_\ell$};
  \node[plus] (aOutSum) at (0,5.51) {$+$};
  \node[state] at (0,.20) {$h_\ell$};
  \draw[flow] (0,.32) -- (aA.south);
  \draw[pathline] (0,.47) -- (1.38,.47);
  \fill[moeink] (0,.47) circle (.034);
  \draw[bypass] (1.38,.47) -- (1.38,1.70) -- (aAttSum.east);
  \draw[flow] (aA.north) -- (aAttSum.south);
  \draw[flow] (aAttSum.north) -- (aR.south);
  \node[state,anchor=west] at ($(aAttSum.north east)+(.15,.08)$) {$z_\ell$};
  \draw[flow] (aR.north) -- (aD.south);
  \moefanout{aD}{a}
  \moefanin{a}{aC}
  \draw[flow] (aC.north) -- (aOutSum.south);
  \draw[bypass] (aAttSum.west) -- (-2.22,1.70) -- (-2.22,5.51) -- (aOutSum.west);
  \draw[flow] (aOutSum.north) -- (0,6.02);
  \node[state,anchor=west] at (.12,6.03) {$h_{\ell+1}$};
\end{scope}

\begin{scope}[shift={(8.17,7.52)}]
  \moepanel{b}{Parallel MoE}
  \node[comp] (bA) at (-1.37,2.09) {$A_\ell$};
  \node[comp] (bR) at (1.03,2.09) {$R_\ell$};
  \node[comm] (bD) at (1.03,2.94) {$D_\ell$};
  \moeexperts{b}{.88}{3.76}
  \node[comm] (bC) at (1.03,4.62) {$C_\ell$};
  \node[plus] (bSum) at (-1.37,5.50) {$+$};
  \node[state] at (0,.20) {$h_\ell$};
  \draw[pathline] (0,.35) -- (0,.72);
  \fill[moeink] (0,.72) circle (.034);
  \draw[flow] (0,.72) -- (-1.37,.72) -- (bA.south);
  \draw[flow] (0,.72) -- (1.03,.72) -- (bR.south);
  \draw[flow] (bR.north) -- (bD.south);
  \moefanout{bD}{b}
  \moefanin{b}{bC}
  \draw[flow] (bA.north) -- (bSum.south);
  \draw[bypass] (-1.37,.72) -- (-2.28,.72) -- (-2.28,5.50) -- (bSum.west);
  \draw[bypass] (bC.north) -- (1.03,5.50) -- (bSum.east);
  \draw[flow] (bSum.north) -- (-1.37,6.02);
  \node[state,anchor=west] at (-1.22,6.03) {$h_{\ell+1}$};
\end{scope}

\begin{scope}[shift={(2.65,0)}]
  \moepanel{c}{Ladder MoE}
  \node[comp] (cR) at (-1.27,1.34) {$R_\ell$};
  \node[comm] (cD) at (-1.27,2.22) {$D_\ell$};
  \moeexperts{c}{-1.37}{3.10}
  \node[comm] (cC) at (-1.27,3.98) {$C_\ell$};
  \node[plus] (cXSum) at (-1.27,5.29) {$+$};
  \node[comp] (cA) at (1.34,2.62) {$A_\ell$};
  \node[plus] (cHSum) at (1.34,4.37) {$+$};
  \node[state] at (-1.27,.19) {$y_\ell$};
  \node[state] at (1.34,.19) {$h_\ell$};
  \draw[pathline] (-1.27,.35) -- (-1.27,.55);
  \fill[moeink] (-1.27,.55) circle (.036);
  \draw[flow] (-1.27,.55) -- (cR.south);
  \draw[bypass] (-1.27,.55) -- (.25,.55) -- (.25,4.37) -- (cHSum.west);
  \draw[flow] (cR.north) -- (cD.south);
  \moefanout{cD}{c}
  \moefanin{c}{cC}
  \draw[flow] (cC.north) -- (cXSum.south);
  \draw[flow] (1.34,.35) -- (cA.south);
  \draw[flow] (cA.north) -- (cHSum.south);
  \draw[pathline] (cHSum.north) -- (1.34,5.29);
  \fill[moeink] (1.34,5.29) circle (.036);
  \draw[flow] (1.34,5.29) -- (1.34,6.02);
  \draw[flow] (1.34,5.29) -- (cXSum.east);
  \draw[flow] (cXSum.north) -- (-1.27,6.02);
  \node[state,anchor=east] at (-1.40,6.06) {$y_{\ell+1}$};
  \node[state,anchor=west] at (1.46,6.06) {$h_{\ell+1}$};
\end{scope}

\begin{scope}[shift={(8.17,0)}]
  \moepanel{d}{Early-Routing MoE ($p=1$)}
  \node[comp] (dA) at (.64,1.02) {$A_\ell$};
  \node[plus] (dAttSum) at (.64,1.74) {$+$};
  \node[comp] (dR) at (-1.42,1.02) {$R_\ell$};
  \node[comm] (dD) at (.64,2.88) {$D_\ell$};
  \moeexperts{d}{.51}{3.73}
  \node[comm] (dC) at (.64,4.56) {$C_\ell$};
  \node[plus] (dSum) at (.64,5.49) {$+$};
  \node[state] at (-1.42,.20) {$z_{\ell-1}$};
  \node[state] at (.64,.20) {$h_\ell$};
  \draw[flow] (-1.42,.36) -- (dR.south);
  \draw[pathline] (.64,.37) -- (.64,.54);
  \fill[moeink] (.64,.54) circle (.034);
  \draw[flow] (.64,.54) -- (dA.south);
  \draw[bypass] (.64,.54) -- (1.62,.54) -- (1.62,1.74) -- (dAttSum.east);
  \draw[flow] (dA.north) -- (dAttSum.south);
  \draw[flow] (dAttSum.north) -- (dD.south);
  \node[state,anchor=west] at ($(dAttSum.north east)+(.15,.08)$) {$z_\ell$};
  \draw[bypass] (dR.north) -- (-1.42,2.47) -- (-.14,2.47) -- (-.14,2.88) -- (dD.west);
  \moefanout{dD}{d}
  \moefanin{d}{dC}
  \draw[flow] (dC.north) -- (dSum.south);
  \draw[bypass] (dAttSum.west) -- (-2.25,1.74) -- (-2.25,5.49) -- (dSum.west);
  \draw[flow] (dSum.north) -- (.64,6.02);
  \node[state,anchor=west] at (.79,6.03) {$h_{\ell+1}$};
\end{scope}

\begin{scope}[shift={(15.49,0)}]
  \draw[frame] (-4.35,-.70) rectangle (4.35,13.92);
  \node[font=\small\bfseries,text=moeink] at (0,-.40) {(e)\enspace BRAID-MoE};

  \node[state] at (-3.17,.18) {$h_\ell$};
  \node[state] at (2.13,.18) {$X_\ell$};

  \node[comp] (eA) at (-3.17,7.40) {$A_\ell$};
  \node[comp] (eR) at (-1.16,7.40) {$R_\ell$};
  \node[comm] (eC) at (2.13,3.23) {$C_{\ell-1}$};
  \node[plus] (eY) at (2.13,5.20) {$+$};
  \node[comm] (eD) at (2.13,7.40) {$D_\ell$};
  \node[plus] (eOut) at (-2.60,11.86) {$+$};

  \node[expert,minimum width=.77cm] (eEone) at (.82,10.04) {$\mathcal{E}_1$};
  \node[expert,minimum width=.77cm] (eEtwo) at (2.13,10.04) {$\mathcal{E}_2$};
  \node[expert,minimum width=.77cm] (eEk) at (3.69,10.04) {$\mathcal{E}_E$};
  \moedots{eEtwo}{eEk}
  \coordinate (eJoin) at (2.13,2.53);
  \draw[pathline] (.72,.55) -- (.72,1.07);
  \draw[flow] (.72,1.07) -- (.72,1.65);
  \draw[pathline] (.72,1.65) -- (.72,1.80) -- (eJoin);
  \draw[pathline] (2.13,.55) -- (2.13,1.07);
  \draw[flow] (2.13,1.07) -- (2.13,1.65);
  \draw[pathline] (2.13,1.65) -- (eJoin);
  \draw[pathline] (3.55,.55) -- (3.55,1.07);
  \draw[flow] (3.55,1.07) -- (3.55,1.65);
  \draw[pathline] (3.55,1.65) -- (3.55,1.80) -- (eJoin);
  \fill[moeink] (eJoin) circle (.040);
  \draw[flow] (eJoin) -- (eC.south);
  \draw[flow] (eC.north) -- (eY.south);

  \draw[pathline] (-3.17,.52) -- (-3.17,5.20);
  \fill[moeink] (-3.17,5.20) circle (.035);
  \draw[flow] (-3.17,5.20) -- (eY.west);
  \draw[flow] (-3.17,5.20) -- (eA.south);
  \draw[flow] (-1.16,5.20) -- (eR.south);
  \fill[moeink] (-1.16,5.20) circle (.034);

  \draw[pathline] (eY.north) -- (2.13,6.17);
  \node[state,anchor=west] at (2.33,5.67) {$y_\ell$};
  \fill[moeink] (2.13,6.17) circle (.038);
  \draw[flow] (2.13,6.17) -- (eD.south);
  \draw[flow] (eR.east) -- (eD.west);
  \node[state,anchor=south] at (.42,7.46) {$\pi_\ell$};

  \draw[white,line width=3.1pt] (-1.16,6.04) -- (-1.16,6.30);
  \draw[pathline] (2.13,6.17) -- (-2.00,6.17) -- (-2.00,11.86);
  \draw[flow] (-2.00,11.86) -- (eOut.east);
  \draw[flow] (eA.north) -- (-3.17,11.86) -- (eOut.west);
  \draw[flow] (eOut.north) -- (-2.60,13.19);
  \node[state,anchor=west] at (-2.46,13.19) {$h_{\ell+1}$};

  \draw[fan] (eD.north) -- (eEone.south);
  \draw[fan] (eD.north) -- (eEtwo.south);
  \draw[fan] (eD.north) -- (eEk.south);
  \draw[flow] (eEone.north) -- (.82,12.96);
  \draw[flow] (eEtwo.north) -- (2.13,12.96);
  \draw[flow] (eEk.north) -- (3.69,12.96);
  \node[state] at (2.13,13.35) {$X_{\ell+1}$};

  \draw[commregion] (1.25,2.88) rectangle (3.05,7.79);
\end{scope}
\end{tikzpicture}
\endgroup%
    }
\caption{Comparison of (a) Sequential, (b) Parallel, (c) Ladder, (d) Early-Routing ($p=1$), and (e) BRAID MoE architectures. Red blocks denote computation and blue blocks denote expert-parallel communication. In BRAID, sharded expert outputs are combined with the residual stream and immediately dispatched to the next layer's experts, enabling back-to-back communication operations.}
    \label{fig:architectures}
\end{figure*}

\section{Our Method: BRAID}
\label{sec:braid}

Building on the complementary ideas of Ladder MoEs and Early-Routing MoEs, we introduce BRAID-MoE, a novel MoE architecture that pairs laddered residual connections with early routing to expose additional communication--computation overlap. The key idea is to decouple routing from the preceding MoE combine while preserving fresh activations for expert computation. This removes attention and routing from the combine-to-dispatch critical path, enabling back-to-back EP communication operations that can overlap with attention.

\subsection{BRAID Architecture}
\label{sec:braid-architecture}

As illustrated in Figure~\ref{fig:architectures}, BRAID-MoE retains the two interleaved residual streams of Ladder MoEs but computes routing decisions from the earlier residual state $h_\ell$, rather than $y_\ell$. The resulting architecture is defined as
\begin{equation}
\begin{aligned}
(S_\ell,\pi_\ell)
    &= \operatorname{Route}_\ell(h_\ell),\\
h_{\ell+1}
    &= y_\ell + \operatorname{Attn}_\ell(h_\ell),\\
y_{\ell+1}
    &= h_{\ell+1} + \operatorname{MoE}_\ell(y_\ell;S_\ell,\pi_\ell).
\end{aligned}
\label{eq:braid}
\end{equation}
In the distributed implementation, $y_\ell$ is obtained by combining the sharded expert outputs of layer $\ell-1$ and adding them to $h_\ell$. Since $h_\ell$ is available before this combine completes, routing is independent of the preceding MoE combine. Experts still process $y_\ell$, which incorporates the preceding MoE output but not the current attention update. Consequently, routing can be performed in advance without requiring experts to process stale activations. The resulting execution schedule is detailed in Algorithm~\ref{alg:braid}.

Unlike Ladder MoEs, BRAID-MoE therefore allows the preceding combine and current dispatch to be scheduled consecutively, with only the residual update separating them. This also creates an opportunity for future fusion of the two communication phases across block boundaries.

\subsection{Multi-Stream Execution}
\label{sec:braid-schedule}

\begin{figure*}[t]
    \centering
    \resizebox{0.95\textwidth}{!}{%
\begingroup
\definecolor{moeink}{HTML}{283849}
\definecolor{moemuted}{HTML}{66798B}
\definecolor{moeblue}{HTML}{397C9D}
\definecolor{moebluefill}{HTML}{E8F2F7}
\definecolor{moered}{HTML}{B65F69}
\definecolor{moeredfill}{HTML}{FBEFF0}
\begin{tikzpicture}[
 x=1cm,y=1cm,font=\small,
 >={Latex[length=1.6mm,width=1.05mm]},
 line cap=round,line join=round,
 flow/.style={draw=moeink,line width=.82pt,-{Latex[length=1.6mm,width=1.05mm]}},
 path/.style={draw=moeink,line width=.82pt},
 blueflow/.style={draw=moeblue,line width=.92pt,-{Latex[length=1.6mm,width=1.05mm]}},
 comm/.style={rectangle,rounded corners=2pt,draw=moeblue,fill=moebluefill,
    line width=.85pt,minimum height=.69cm,minimum width=1.52cm,inner sep=3pt,align=center},
 norm/.style={comm,minimum width=1.75cm},
 comp/.style={rectangle,rounded corners=2pt,draw=moered,fill=moeredfill,
    line width=.85pt,minimum height=.56cm,minimum width=.78cm,inner sep=2pt,align=center},
 add/.style={circle,draw=moeink,fill=white,line width=.82pt,minimum size=.48cm,inner sep=0pt},
 state/.style={font=\small,text=moeink,inner sep=1pt},
 muted/.style={font=\footnotesize,text=moemuted,inner sep=1pt},
 commregion/.style={draw=moeblue!55,densely dashed,line width=.78pt,rounded corners=5pt}
]
\node[comm] (C) at (4.55,3.95) {$C_{\ell-1}$};
\node[add] (plus) at (6.54,3.95) {$+$};
\node[norm] (N) at (8.90,3.95) {RMSNorm};
\node[comm] (D) at (11.40,3.95) {$D_\ell$};
\draw[commregion] (3.48,3.42) rectangle (12.47,4.50);
\node[state,anchor=east] at (.90,4.89) {$X_{\ell,1}$};
\node[state,anchor=east] at (.90,3.95) {$X_{\ell,2}$};
\node[state,anchor=east] at (.90,3.01) {$X_{\ell,E}$};
\foreach \yy in {3.31,3.48,3.65} {\fill[black] (1.12,\yy) circle (.026cm);} 
\draw[flow] (1.04,4.89) -- (3.12,4.89) -- ([yshift=+.20cm]C.west);
\draw[flow] (1.04,3.95) -- ([yshift=0cm]C.west);
\draw[flow] (1.04,3.01) -- (3.12,3.01) -- ([yshift=-.20cm]C.west);
\draw[flow] (C.east) -- (plus.west);
\coordinate (ybranch) at (7.40,3.95);
\draw[path] (plus.east) -- (ybranch);
\fill[moeink] (ybranch) circle (.046cm);
\draw[flow] (ybranch) -- (N.west);
\draw[flow] (N.east) -- (D.west);
\node[state,anchor=east] at (.90,2.13) {$h_\ell$};
\draw[flow] (1.05,2.13) -- (6.54,2.13) -- (plus.south);
\node[state,anchor=east] at (.90,5.77) {$\pi_\ell$};
\draw[flow] (1.05,5.77) -- (11.40,5.77) -- (D.north);
\draw[flow] (ybranch) -- (7.40,2.13) -- (15.48,2.13);
\node[state,anchor=west] at (15.60,2.13) {$y_\ell$};
\node[comp] (E1) at (13.75,4.89) {$\mathcal{E}_1$};
\node[comp] (E2) at (13.75,3.95) {$\mathcal{E}_2$};
\node[comp] (EE) at (13.75,3.01) {$\mathcal{E}_E$};
\foreach \yy in {3.31,3.48,3.65} {\fill[black] (14.80,\yy) circle (.026cm);}  
\draw[flow] ([yshift=+.20cm]D.east) -- (12.55,4.15) -- (13.36,4.60);
\draw[flow] (D.east) -- (13.36,3.95);
\draw[flow] ([yshift=-.20cm]D.east) -- (12.55,3.75) -- (13.36,3.30);
\draw[flow] (E1.east) -- (15.48,4.89);
\draw[flow] (E2.east) -- (15.48,3.95);
\draw[flow] (EE.east) -- (15.48,3.01);
\node[state,anchor=west] at (15.60,4.89) {$X_{\ell+1,1}$};
\node[state,anchor=west] at (15.60,3.95) {$X_{\ell+1,2}$};
\node[state,anchor=west] at (15.60,3.01) {$X_{\ell+1,E}$};
\end{tikzpicture}
\endgroup
    }
\caption{Communication dataflow in BRAID-MoE. The dashed region highlights the back-to-back combine and dispatch operations enabled by precomputed routing assignments. Sharded expert outputs are combined with the residual stream to produce $y_\ell$. In our implementation, residual addition and MoE RMSNorm are fused into the combine epilogue, with normalization applied only on the dispatch path. The unnormalized residual $y_\ell$ is preserved for the subsequent residual update.}
    \label{fig:comm-dataflow}
\end{figure*}
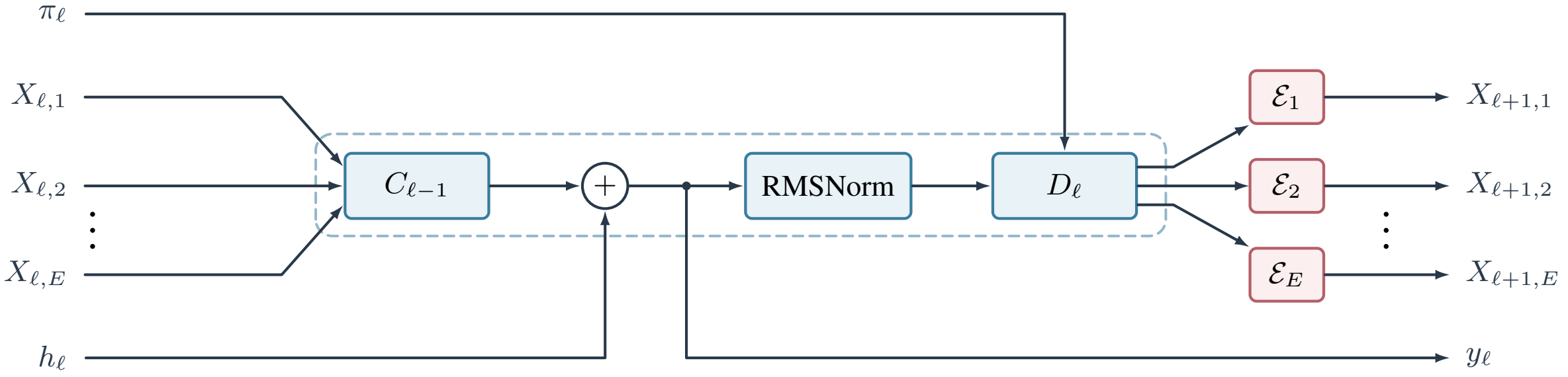

\paragraph{Consecutive communication kernels.}
As illustrated in Figure~\ref{fig:architectures}, the communication stream first combines the sharded expert outputs $X_\ell$ of layer $\ell-1$ with the residual $h_\ell$ to produce $y_\ell$, which is then immediately dispatched to the next experts using the precomputed routing assignments $\pi_\ell$. In our implementation, we extend DeepEP's \texttt{EPBuffer} combine kernel by fusing the residual addition and the MoE-specific RMSNorm into its epilogue, while preserving the unnormalized residual $y_\ell$ for the subsequent residual update, as detailed in Figure~\ref{fig:comm-dataflow}. This avoids a separate normalization kernel and reduces intermediate memory traffic. Importantly, combine and dispatch remain distinct communication kernels.

\paragraph{Overlapping communication and computation.} 
We translate BRAID-MoE's architectural properties into a two-stream execution schedule, illustrated alongside those of Sequential, Parallel, Ladder, and Early-Routing MoEs in Figure~\ref{fig:schedules}. We dedicate one CUDA stream to communication ($C,D$) and another to computation ($R,A,\mathcal{E}$), using CUDA events to synchronize only when required by data dependencies. Since routing and attention both consume $h_\ell$, they can execute independently of the preceding combine operation. Routing can therefore overlap with combine, while attention can overlap with both combine and dispatch. Consequently, the consecutive communication operations can be largely hidden behind routing and attention computation.

The remaining dependency concerns expert computation, which must wait for dispatch to complete before consuming the redistributed activations. Similarly, the following combine cannot begin before the corresponding expert outputs are ready. Thus, unlike routing and attention, expert computation cannot be hidden behind the communication operations of its own layer in this schedule. Nevertheless, its execution may overlap with independent communication from other operations, such as tensor-parallel collectives in subsequent attention layers, provided their dependencies and scheduling permit.

\paragraph{Implementation considerations.} Although separating communication and computation into different CUDA streams exposes concurrency, actual overlap depends on the availability of GPU resources. In particular, DeepEP communication kernels consume SM resources, potentially competing with routing and attention computations. Allocating more SMs to communication can accelerate individual collectives but reduce the resources available for concurrent computation. Conversely, limiting communication SM occupancy may improve overlap at the expense of standalone communication throughput. BRAID therefore introduces a resource-allocation trade-off: the objective is not necessarily to maximize the throughput of each kernel independently, but to minimize end-to-end layer latency by balancing communication progress against computation throughput.

\begin{figure*}[t]
    \centering
    \resizebox{\textwidth}{!}{%
\begingroup
\definecolor{schedink}{HTML}{26354B}
\definecolor{schedmuted}{HTML}{607188}
\definecolor{schedred}{HTML}{B65C69}
\definecolor{schedredfill}{HTML}{FAEAEC}
\definecolor{schedblue}{HTML}{3C83A2}
\definecolor{schedbluefill}{HTML}{E7F2F8}
\begin{tikzpicture}[
  x=.56cm,y=.84cm,
  >=Latex,
  line cap=round,line join=round,
  cpubox/.style={draw=schedred,fill=schedredfill,line width=.65pt,rounded corners=1.6pt},
  commbox/.style={draw=schedblue,fill=schedbluefill,line width=.65pt,rounded corners=1.6pt}
]
  \newcommand{\lanes}[3][11.7]{%
    \node[font=\small\bfseries,text=schedink,anchor=center]
       at ({#1/2},2.16) {(#2)\enspace #3};
    \fill[black!2.8] (0,.89) rectangle (#1,1.65);
    \fill[black!2.8] (0,-.05) rectangle (#1,.71);
    \draw[black!10,line width=.35pt] (0,.89) rectangle (#1,1.65);
    \draw[black!10,line width=.35pt] (0,-.05) rectangle (#1,.71);
    \node[font=\scriptsize,text=schedmuted,anchor=east,align=right]
      at (-.20,1.27) {Compute\\stream};
    \node[font=\scriptsize,text=schedmuted,anchor=east,align=right]
      at (-.20,.33) {Comm.\\stream};
    \draw[->,black!45,line width=.45pt] (0,-.30) -- ({#1-.15},-.30);
    \node[font=\scriptsize,text=schedmuted,anchor=west]
      at ({#1-.10},-.30) {time};
  }
  \newcommand{\cpu}[3]{%
    \draw[cpubox] (#1,.96) rectangle (#2,1.59);
    \node[font=\scriptsize,text=schedink,inner sep=0pt]
      at ({(#1+#2)/2},1.275) {#3};
  }
  \newcommand{\comm}[3]{%
    \draw[commbox] (#1,.02) rectangle (#2,.65);
    \node[font=\scriptsize,text=schedink,inner sep=0pt]
      at ({(#1+#2)/2},.335) {#3};
  }

  \newcommand{\cpulong}[3]{\cpu{#1}{#2}{\scalebox{.84}{#3}}}
  \newcommand{\commlong}[3]{\comm{#1}{#2}{\scalebox{.84}{#3}}}

  \begin{scope}[shift={(0,0)}]
    \lanes{a}{Sequential MoE}
    \cpu{.15}{1.60}{$A_\ell$}
    \cpu{1.60}{2.70}{$R_\ell$}
    \comm{2.70}{3.70}{$D_\ell$}
    \cpu{3.70}{4.80}{$\mathcal{E}_\ell$}
    \comm{4.80}{5.85}{$C_\ell$}
    \cpulong{5.85}{7.30}{$A_{\ell+1}$}
    \cpulong{7.30}{8.40}{$R_{\ell+1}$}
    \commlong{8.40}{9.40}{$D_{\ell+1}$}
    \cpulong{9.40}{10.50}{$\mathcal{E}_{\ell+1}$}
    \commlong{10.50}{11.55}{$C_{\ell+1}$}
  \end{scope}

  \begin{scope}[shift={(15.3,0)}]
    \lanes{b}{Parallel MoE}
    \cpu{.55}{1.55}{$R_\ell$}
    \cpu{1.55}{3.20}{$A_\ell$}
    \comm{1.55}{2.75}{$D_\ell$}
    \cpu{3.20}{4.50}{$\mathcal{E}_\ell$}
    \comm{4.50}{5.55}{$C_\ell$}
    \cpulong{5.55}{6.55}{$R_{\ell+1}$}
    \cpulong{6.55}{8.20}{$A_{\ell+1}$}
    \commlong{6.55}{7.75}{$D_{\ell+1}$}
    \cpulong{8.20}{9.50}{$\mathcal{E}_{\ell+1}$}
    \commlong{9.50}{10.55}{$C_{\ell+1}$}
  \end{scope}

  \begin{scope}[shift={(0,-2.96)}]
    \lanes{c}{Ladder MoE}
    \commlong{.55}{1.60}{$C_{\ell-1}$}
    \cpu{.55}{1.95}{$A_\ell$}
    \cpu{1.95}{3.00}{$R_\ell$}
    \comm{3.00}{4.00}{$D_\ell$}
    \cpu{4.00}{5.20}{$\mathcal{E}_\ell$}
    \comm{5.20}{6.25}{$C_\ell$}
    \cpulong{5.20}{6.60}{$A_{\ell+1}$}
    \cpulong{6.60}{7.65}{$R_{\ell+1}$}
    \commlong{7.65}{8.65}{$D_{\ell+1}$}
    \cpulong{8.65}{9.85}{$\mathcal{E}_{\ell+1}$}
    \commlong{9.85}{10.90}{$C_{\ell+1}$}
  \end{scope}

  \begin{scope}[shift={(15.3,-2.96)}]
    \lanes{d}{Early-Routing MoE ($p=1$)}
    \commlong{.55}{1.67}{$C_{\ell-1}$}
    \cpu{.55}{1.65}{$R_\ell$}
    \cpu{1.67}{3.04}{$A_\ell$}
    \comm{3.04}{4.15}{$D_\ell$}
    \cpu{4.15}{5.27}{$\mathcal{E}_\ell$}
    \comm{5.27}{6.39}{$C_\ell$}
    \cpulong{5.27}{6.37}{$R_{\ell+1}$}
    \cpulong{6.39}{7.76}{$A_{\ell+1}$}
    \commlong{7.76}{8.87}{$D_{\ell+1}$}
    \cpulong{8.87}{9.99}{$\mathcal{E}_{\ell+1}$}
    \commlong{9.99}{11.11}{$C_{\ell+1}$}
    \cpulong{9.99}{11.09}{$R_{\ell+2}$}
  \end{scope}

  \begin{scope}[shift={(7.65,-5.92)}]
    \lanes[10.0]{e}{BRAID-MoE}
    \commlong{.55}{2.10}{$C_{\ell-1}$}
    \cpu{.55}{1.70}{$R_\ell$}
    \cpu{1.70}{3.65}{$A_\ell$}
    \comm{2.10}{3.65}{$D_\ell$}
    \cpu{3.65}{5.15}{$\mathcal{E}_\ell$}
    \comm{5.15}{6.70}{$C_\ell$}
    \cpulong{5.15}{6.30}{$R_{\ell+1}$}
    \cpulong{6.30}{8.25}{$A_{\ell+1}$}
    \commlong{6.70}{8.25}{$D_{\ell+1}$}
    \cpulong{8.25}{9.75}{$\mathcal{E}_{\ell+1}$}
  \end{scope}
\end{tikzpicture}
\endgroup%
    }
\caption{Illustrative two-stream execution schedules of (a) Sequential, (b) Parallel, (c) Ladder, (d) Early-Routing ($p=1$), and (e) BRAID MoEs. Red blocks denote routing ($R$), attention ($A$), and expert computation ($\mathcal{E}$), while blue blocks denote dispatch ($D$) and combine communication ($C$). (a) Sequential MoE serializes attention and MoE execution. (b) Parallel MoE overlaps dispatch with attention, since both branches consume the same residual. (c) Ladder MoE overlaps attention with the preceding combine, but routing still depends on that combine. (d) Early-Routing computes routing from an earlier representation, overlapping it with the preceding combine, while dispatch still waits for the fresh post-attention representation. (e) BRAID computes routing from the earlier residual while sharded expert outputs are combined, enabling back-to-back combine and dispatch operations across consecutive layers, with communication overlapping attention. Timings are schematic and not measured.}
    \label{fig:schedules}
\end{figure*}
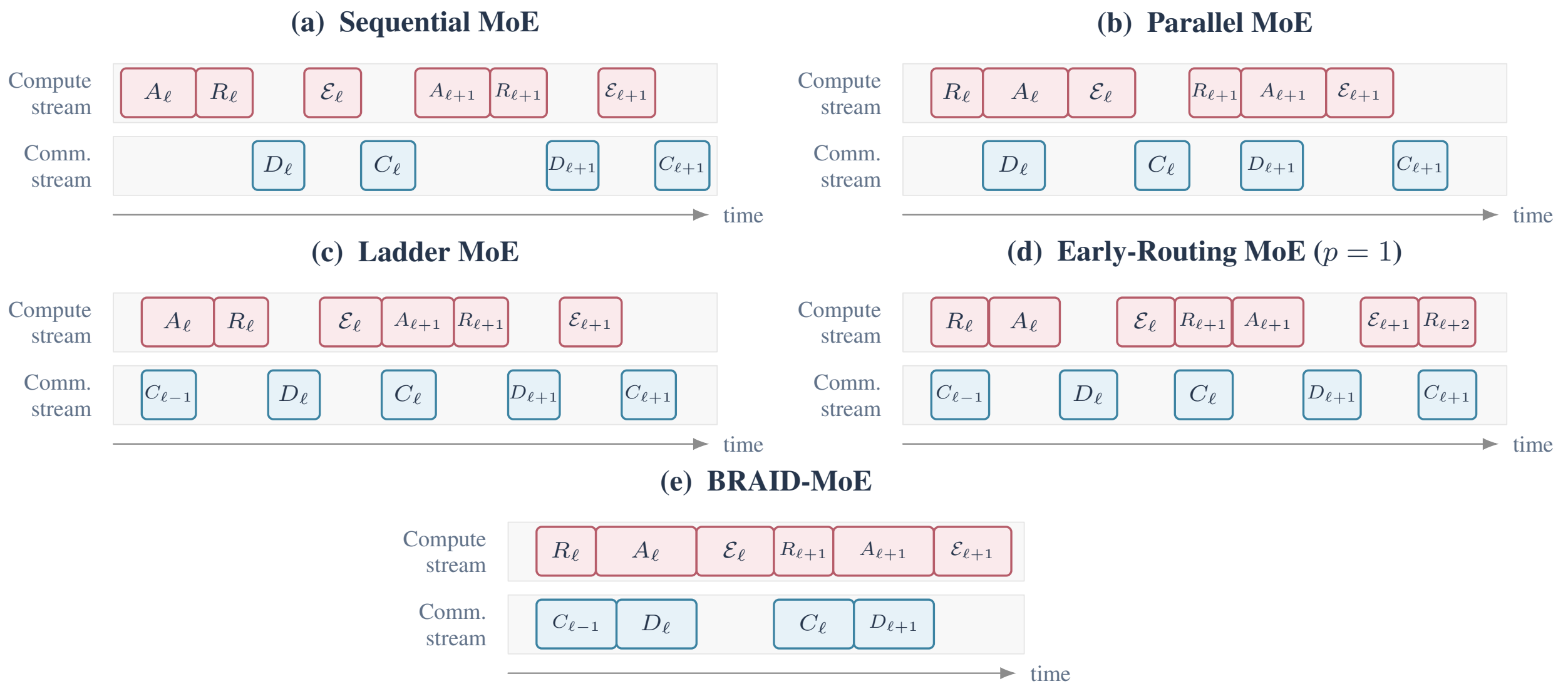

\begin{algorithm}[t]
\caption{BRAID-MoE execution at layer $\ell$. Here, \textsc{Record} marks an event on the current stream, and \textsc{Wait} inserts a stream-local dependency on an event produced by the other stream. These waits do not block the CPU.}
\label{alg:braid}
\begin{algorithmic}[1]
\Require Residual $h_\ell$, sharded expert outputs $X_\ell$, retained combine metadata $\mu_{\ell-1}$
\Ensure Residual $h_{\ell+1}$, sharded expert outputs $X_{\ell+1}$, retained metadata $\mu_\ell$

\Statex \textbf{Computation stream} $\mathcal{S}_{\mathrm{comp}}$
\State $(S_\ell,\pi_\ell) \gets \operatorname{Route}_\ell(h_\ell)$
\State \Call{Record}{$e_\ell^{R}$}
\State $a_\ell \gets \operatorname{Attn}_\ell(h_\ell)$
\State \Call{Wait}{$e_\ell^{Y}$}
\State $h_{\ell+1} \gets y_\ell + a_\ell$
\State \Call{Wait}{$e_\ell^{D}$}
\State $X_{\ell+1} \gets \operatorname{Expert}_\ell(v_\ell)$
\State \Call{Record}{$e_\ell^{E}$}

\Statex
\Statex \textbf{Communication stream} $\mathcal{S}_{\mathrm{comm}}$
\State \Call{Wait}{$e_{\ell-1}^{E}$}
\State $m_\ell \gets \operatorname{Combine}_{\ell-1}(X_\ell;\mu_{\ell-1})$
\State $y_\ell \gets h_\ell + m_\ell$
\State \Call{Record}{$e_\ell^{Y}$}
\State \Call{Wait}{$e_\ell^{R}$}
\State $(v_\ell,\mu_\ell) \gets \operatorname{Dispatch}_\ell(y_\ell;S_\ell,\pi_\ell)$
\State \Call{Record}{$e_\ell^{D}$}

\end{algorithmic}
\end{algorithm}

\section{Training Experiments}
\label{sec:quality}

As in \citet{zhang2025ladder}, our training experiments pursue three objectives: (i) to evaluate the scalability of BRAID when pretrained from scratch, (ii) to assess whether an existing pretrained MoE model can be converted into a BRAID architecture through a limited amount of continued training, while preserving its original model quality, and (iii) to investigate whether supervised fine-tuning can recover downstream performance after converting an instruction-tuned MoE to BRAID.

\subsection{Pretraining from Scratch}
\label{sec:pretraining}

\paragraph{Experimental setup.}
We consider Qwen3-inspired MoE architectures~\citep{qwen2025qwen3} at two model scales. The smaller model has 99.8M active and 392M total parameters, while the larger model has 504M active and 2.51B total parameters. Both models use $E=64$ routed experts, top-$K$ routing with $K=8$, and no shared experts. Following standard practice, attention and expert matrices are optimized with Muon~\citep{jordan2024muon}, while embeddings, routers, normalization parameters, and the language-model head use AdamW~\citep{loshchilov2019decoupled}. We use a 5\% linear warmup followed by cosine decay to 10\% of the peak learning rate. The global batch contains 512 sequences of length 2,048, corresponding to 1,048,576 tokens per optimization step at both scales. Both models are pretrained on FineWeb v1.4.0~\citep{penedo2024fineweb} using the Llama-2 tokenizer~\citep{touvron2023llama2} with a vocabulary of 32,000 tokens. Evaluation uses a fixed held-out stream containing 100,139,008 tokens with no overlap with the training data set. We use the Joint MoE scaling-law prescription~\citep{ludziejewski2025joint} to determine the training budgets, obtaining 3.27B tokens for the 0.4B-A0.1B model and 10.46B tokens for the 2.5B-A0.5B model. Appendix~\ref{app:model-token-allocation} details the model--token allocation rationale, its local IsoFLOP validation, and the data streams used.

We first train the sequential MoE baseline and subsequently train three architectural variants: Parallel MoE, Ladder MoE, and BRAID-MoE. BRAID uses a separate RMSNorm for its early router, distinct from the expert-input normalization. Importantly, \emph{we keep the entire training recipe unchanged across architectures}, including the optimizer, learning-rate schedule, batch size, training duration, and all other hyperparameters. No architecture-specific tuning is performed, allowing us to isolate the effect of the architectural modifications on training dynamics and final model quality. Further architecture, initialization, and optimization details are provided in Appendix~\ref{app:details}; hardware specifications are given in Appendix~\ref{app:hardware}.

\begin{figure}[t]
    \centering
    \includegraphics[width=\linewidth]{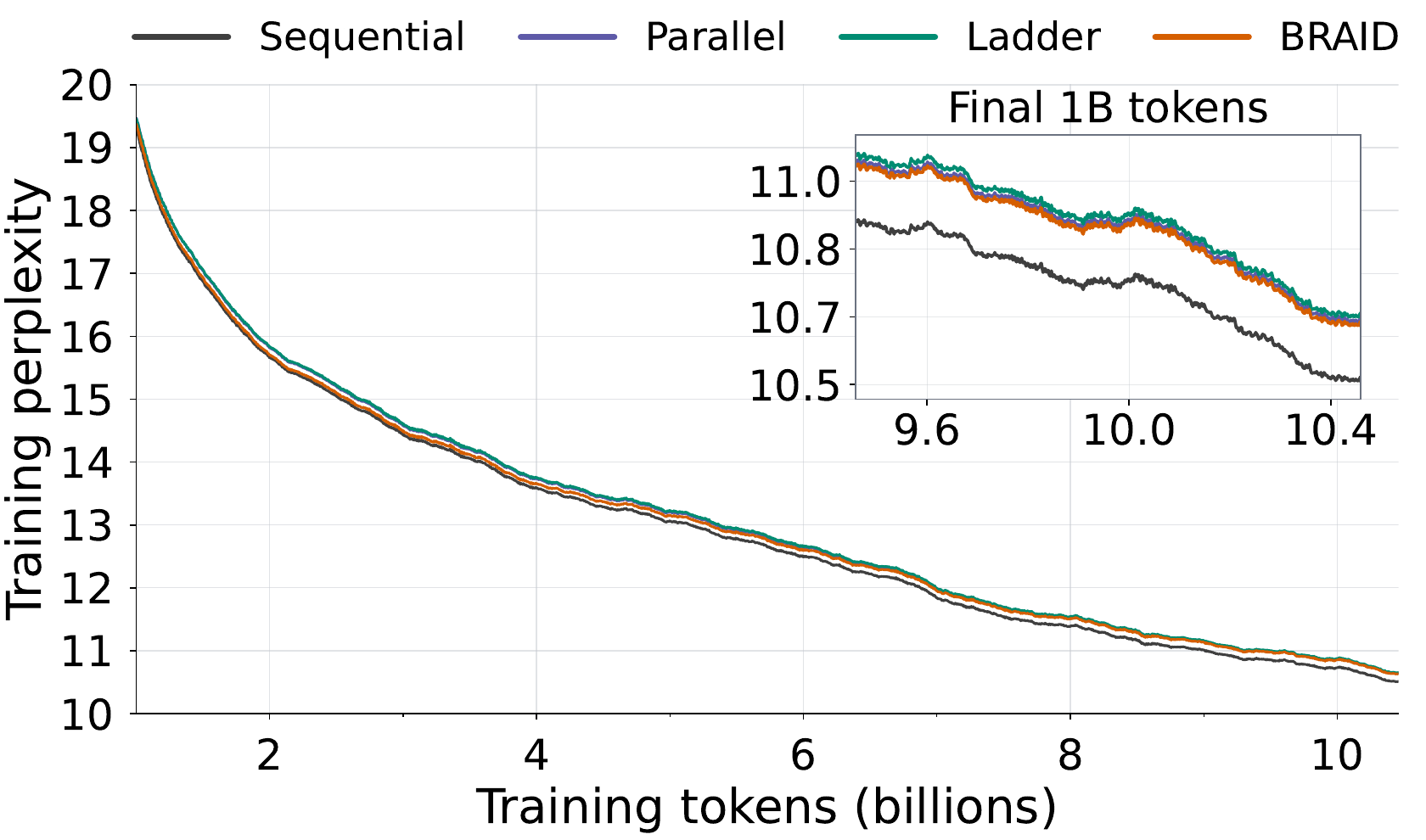}
    \caption{Training perplexity for the 2.5B-A0.5B Qwen3-style MoE under Sequential, Parallel, Ladder, and BRAID schedules. Curves are smoothed over a trailing window of 0.3B training tokens. The main panel covers 1B--10.46B tokens, while the inset enlarges the final 1B tokens. All schedules use matched training settings. BRAID uses a separate normalization layer for early routing.}
    \label{fig:pretraining-ppl}
\end{figure}

\paragraph{Results.} Figure~\ref{fig:pretraining-ppl} reports the 2.5B-A0.5B training perplexity as a function of the number of tokens processed. To facilitate a controlled comparison, all experiments use the same data loader seed. We observe that all three alternative architectures exhibit stable convergence, with training trajectories closely matching that of the sequential MoE baseline. Parallel MoE, plain Ladder MoE, and BRAID-MoE achieve similar training perplexities.

Table~\ref{tab:pretrain} reports the final validation losses for both model sizes. For model size 0.4B-A0.1B, the three alternative architectures remain within 1.02\% of the sequential baseline. At 2.5B-A0.5B, all three alternative architectures remain within 0.6\% of the sequential baseline: Parallel, Ladder, and BRAID differ by 0.53\%, 0.58\%, and 0.49\%, respectively. These results suggest that the additional parallelism introduced by the alternative schedules can be achieved without materially compromising model quality, even when using an unchanged training recipe.

Finally, we emphasize that our comparison fixes the number of training tokens rather than the wall-clock training time. This protocol isolates model quality from differences in execution efficiency. An alternative comparison under a fixed wall-clock budget could favor architectures that support greater communication--computation overlap, allowing more training tokens to be processed within the same time budget. Evaluating this potential advantage requires dedicated end-to-end training throughput measurements.

\begin{table}[t]
\centering\small
\caption{Pretraining comparisons at a fixed token budget. Loss is held-out validation cross-entropy. Differences are relative to the Sequential schedule. Relative differences are computed from unrounded cross-entropies.}
\label{tab:pretrain}
\setlength{\tabcolsep}{2pt}
\begin{tabular*}{\linewidth}{@{\extracolsep{\fill}}llrr@{}}
\toprule
Scale & Schedule &
\shortstack{Validation\\cross-entropy} &
\shortstack{Difference vs.\\Sequential} \\
\midrule
\multirow{4}{58pt}{\shortstack{0.4B-A0.1B\\3.27B tokens}}
& Sequential   & 2.744 & -- \\
& Parallel     & \textbf{2.770} & \textbf{+0.94\%} \\
& Ladder & 2.772 & +1.02\% \\
& BRAID        & 2.771 & +0.98\% \\
\midrule
\multirow{4}{58pt}{\shortstack{2.5B-A0.5B\\10.46B tokens}}
& Sequential   & 2.345 & -- \\
& Parallel     & 2.357 & +0.53\% \\
& Ladder       & 2.358 & +0.58\% \\
& BRAID        & \textbf{2.356} & \textbf{+0.49\%} \\
\bottomrule
\end{tabular*}
\end{table}

\subsection{Adapting Sequential Checkpoints}
\label{sec:checkpoint-adaptation}

\paragraph{Problem setting.}
We investigate whether a pretrained sequential MoE can be efficiently converted into BRAID through a short period of continued training, rather than requiring pretraining from scratch. We initialize BRAID from the final checkpoint of a pretrained sequential MoE and continue training on a FineWeb v1.4.0 data stream disjoint from pretraining and validation. We convert the final half of the transformer layers of each model to the BRAID architecture, corresponding to layers 7 to 13 for 0.4B-A0.1B and layers 12 to 23 for 2.5B-A0.5B. Existing model weights are copied directly, and each added early router RMSNorm is initialized from the corresponding trained expert-input RMSNorm.

As a control, we continue training the original sequential architecture at the same continuation token budget. Both architectures use the same optimizer groups and learning-rate schedule shape, but with independently tuned peak learning rates. Further continuation hyperparameters and data-stream details are provided in Appendix~\ref{app:checkpoint-continuation}.

\paragraph{Recovery.} We limit continued training to approximately 10\% of the original pretraining budget, corresponding to 313 additional steps for 0.4B-A0.1B and 998 steps for 2.5B-A0.5B. Despite this modest additional compute, BRAID recovers performance close to the original sequential checkpoints, with held-out validation cross-entropy within 0.9\% for 0.4B-A0.1B and 0.3\% for 2.5B-A0.5B of the respective sequential checkpoints before continuation, as shown in Table~\ref{tab:adapt}. These results demonstrate that existing sequential MoEs can be adapted to BRAID with limited additional training rather than requiring pretraining from scratch.

\begin{table}[t]
\centering\small
\caption{Held-out validation cross-entropy before and after checkpoint continuation. Continuation uses 313 steps at 0.4B-A0.1B and 998 steps at 2.5B-A0.5B. The sequential and BRAID continuations use $0.1\times$ and $0.3\times$ the pretraining peak learning rate, respectively.}
\label{tab:adapt}
\setlength{\tabcolsep}{3pt}
\begin{tabular*}{\linewidth}{@{\extracolsep{\fill}}lrrrr@{}}
\toprule
& \multicolumn{2}{c}{Sequential}
& \multicolumn{2}{c}{BRAID} \\
\cmidrule(lr){2-3}
\cmidrule(lr){4-5}
Scale & Before & After & Before & After \\
\midrule
0.4B-A0.1B & 2.744 & 2.726 & 2.932 & 2.768 \\
2.5B-A0.5B & 2.345 & 2.319 & 2.416 & 2.352 \\
\bottomrule
\end{tabular*}
\end{table}

\subsection{Post-Training Recovery}
\label{sec:post-training-recovery}

\paragraph{Conversion and recovery protocol.}
We further investigate whether supervised fine-tuning (SFT) can recover downstream task performance after converting a pretrained sequential MoE into BRAID. To this end, we use OLMoE-1B-7B-0125-Instruct~\citep{muennighoff2025olmoe} and convert the upper half of its Transformer layers to BRAID, leaving the remaining layers unchanged, as in \citet{zhang2025ladder}. We evaluate three configurations: (i) the original sequential model, (ii) the converted BRAID model before SFT, and (iii) the same model after a short SFT recovery phase. The conversion follows the initialization procedure described in Section~\ref{sec:checkpoint-adaptation}, and all model parameters are updated during recovery. We fine-tune the converted model for two epochs on a 5\% sample of the Infinity-Instruct 7M and Gen subsets%
\footnote{\url{https://huggingface.co/datasets/BAAI/Infinity-Instruct}},
corresponding to approximately 205.8M tokens per epoch. Only assistant tokens contribute to the training loss. We use the Open Instruct framework%
\footnote{\url{https://github.com/allenai/open-instruct}} with AdamW, a peak learning rate of $5\times10^{-6}$, 200 linear warmup updates followed by cosine decay to zero, and a global batch size of 32 conversations. This deliberately limited training budget allows us to assess whether BRAID can be adopted from an existing instruction-tuned checkpoint without requiring extensive additional training. Further optimization and implementation details are provided in Appendix~\ref{app:olmoe-sft}. 

\paragraph{Recovery results.}
We evaluate downstream performance using OLMES%
\footnote{\url{https://github.com/allenai/olmes}}~\citep{gu2025olmes}
on its nine core tasks: ARC-Challenge, ARC-Easy, BoolQ, CommonsenseQA, HellaSwag, OpenBookQA, PIQA, Social IQa, and WinoGrande. All three configurations are evaluated under identical conditions, including task prompts, formatting, few-shot settings, and metric implementations. Table~\ref{tab:olmoe-sft-recovery} compares downstream accuracy across the three configurations. Converting the upper half of the model to BRAID reduces average accuracy from 74.6 to 60.2, while the subsequent SFT recovery phase brings it to 70.2. The recovered model is within 4.4 percentage points of the original sequential checkpoint. These results provide evidence that BRAID can be introduced into an existing instruction-tuned MoE with a limited additional training budget, without requiring full pretraining from scratch.

\begin{table*}[t]
\centering\small
\caption{Post-training recovery for OLMoE-1B-7B-0125-Instruct (1B active, 7B total parameters) on the nine core OLMES tasks. Each task uses 5-shot evaluation without chat formatting and reports the better of multiple-choice and cloze accuracy, following OLMES. Avg.\ is the unweighted mean across the nine tasks. All values are percentages.}
\label{tab:olmoe-sft-recovery}
\setlength{\tabcolsep}{3pt}
\begin{tabular}{lrrrrrrrrrr}
\toprule
Model & ARC-C & ARC-E & BoolQ & \shortstack{Common-\\senseQA} & HellaSwag & \shortstack{OpenBook\\QA} & PIQA & \shortstack{Social\\IQa} & WinoGrande & Avg. \\
\midrule
Original OLMoE & 68.8 & 84.1 & 81.4 & 74.0 & 82.3 & 65.6 & 80.4 & 66.2 & 68.3 & 74.6 \\
\tabdashline{11}\noalign{\vskip 2pt}
BRAID-half (pre-SFT) & 47.0 & 72.8 & 44.9 & 64.6 & 66.8 & 51.6 & 76.1 & 53.5 & 64.9 & 60.2 \\
BRAID-half + SFT & 58.4 & 82.1 & 81.7 & 71.8 & 73.9 & 59.6 & 77.4 & 62.2 & 64.7 & 70.2 \\
\bottomrule
\end{tabular}
\end{table*}

\section{Inference Experiments}
\label{sec:perf}
We now assess the inference efficiency of our method and compare our four execution schedules: Sequential, Parallel, Ladder, and BRAID.

\subsection{Setup}

\paragraph{Infrastructure.}
Experiments run on four-H100 nodes. Intra-node communication uses NVLink (up to 900\,GB/s aggregate bidirectional bandwidth per GPU), while nodes are interconnected through InfiniBand NDR400 (400\,Gb/s per link). We compare EP=4 within one node and EP=8 across two nodes; further hardware details are given in Appendix~\ref{app:hardware}.

\paragraph{Model and execution engine.}
We benchmark Qwen3-30B-A3B-Base in BF16 with its native tokenizer, using vLLM~\citep{kwon2023efficient} and DeepEP~\citep{deepseek2025deepep}. Benchmark prompts are drawn from a subset of FineWeb~v1.4.0~\citep{penedo2024fineweb}. The architectural variants are not retrained: as a consequence, we replay the same expert assignments across schedules to compare their execution efficiency fairly. We also use the same static expert placement, computed with DeepSeek's expert-parallel load balancer (EPLB)~\citep{deepseekai2025deepseekv3technicalreport} without expert replication, to improve GPU load balance on these benchmark prompts. It is computed once from expert loads on held-out FineWeb tokens and is not rebalanced during inference. Each variant uses its available kernel fusions (\hyperref[par:fusions]{Appendix~\ref*{app:inference-details} (Fusions)}). Decode uses CUDA graph replay to prevent launch overhead from dominating total execution time. 

\paragraph{Communication overlap and resource contention.}
The benefit of communication--computation overlap depends on the amount of independent computation available and the extent of resource contention. Ideally, communication can be hidden behind independent computation, with the maximum overlap limited by the shorter of the communication duration and the available computation window. However, concurrent execution does not inherently improve performance when operations compete for the same bottleneck resource. Communication and computation can overlap efficiently because they primarily rely on different resources: interconnect bandwidth versus compute throughput or HBM bandwidth. Nevertheless, resource contention remains. DeepEP dispatch and combine transfers require a small number of SMs to execute communication kernels, while their epilogues additionally consume SM capacity and memory bandwidth for token copying and expansion during dispatch and reduction during combine~\citep{deepseek2025deepep}. This contention limits the achievable overlap. We therefore tune SM allocations for transfers and epilogues separately for each variant, using held-out prompts (Appendix~\ref{app:tuning}).        

\paragraph{Methodology.}
We measure prefill latency and latency per decode step over 32 generated tokens. Each workload is repeated in five independent jobs of 30 rounds, with all four schedules run in randomized order in each round. We report the mean of job medians $\pm$ standard error. Further details are in Appendix~\ref{app:inference-timing}.

\subsection{Results: Speedups and Overlaps}
Parallel and Ladder primarily overlap dispatch and combine, respectively, with attention and nearby residual/normalization work. BRAID additionally overlaps early routing and schedules combine and dispatch back-to-back, allowing both communication phases to overlap with attention-side computation (Figure~\ref{fig:schedules}).

We first consider the reference workload of 16 prompts of 1,024 tokens per GPU (Table~\ref{tab:perf}). For EP=4, communication remains within a single node over NVLink and is sufficiently short for BRAID to overlap both combine and dispatch extensively with computation. BRAID achieves $1.171\times$ prefill and $1.182\times$ decode speedups over Sequential, overlapping 96.3\%/92.8\% of combine/dispatch in prefill and 89.2\%/82.7\% in decode. This translates into a clear advantage over both Parallel and Ladder.

For EP=8, inter-node communication over InfiniBand is substantially longer than the available computation window. As a result, most of the combine can still overlap with computation (95.3\% in prefill and 88.8\% in decode), but the subsequent dispatch remains largely exposed (31.3\% and 5.8\% overlap, respectively). BRAID nevertheless achieves $1.178\times$ prefill and $1.278\times$ decode speedups over Sequential. Its additional gains over Parallel and Ladder are more limited, however, since these schedules can also overlap one communication phase. In particular, BRAID outperforms Parallel by only 0.3\% in prefill.

Figures~\ref{fig:sweeps} and~\ref{fig:sweeps8} show speedups as batch size $B$ or context length $L$ varies, for EP=4 and EP=8 respectively. Prefill and decode exhibit different trends because their compute and communication costs scale differently. Figure~\ref{fig:vllm-execution-traces} in Appendix~\ref{app:inference} provides representative EP=4 prefill execution traces illustrating the different schedules.

\begin{table*}[t]
\centering\small
\caption{Qwen3-30B-A3B inference for 16 prompts of 1,024 tokens per GPU
(global batch 64 with EP=4 and 128 with EP=8). Latencies are the mean of
five job medians $\pm$ standard error (30 rounds per job, each measuring all four schedules in random order); decode latency is per step. Combine and dispatch overlaps are medians over three Nsight Systems traces of the fraction of each communication kernel's execution time concurrent with computation kernels on the same GPU. Speedup is relative to Sequential; bold marks the best latency and speedup.}
\label{tab:perf}
\setlength{\tabcolsep}{3pt}
\begin{tabular}{llrr!{\tabdashvline}rrrr!{\tabdashvline}rr}
\toprule
& & \multicolumn{4}{c}{Prefill} & \multicolumn{4}{c}{Decode} \\
\cmidrule(lr){3-6}\cmidrule(lr){7-10}
Topology & Schedule & \shortstack{Latency\\(ms)} & \multicolumn{1}{r!{\tabdashvline[15pt]}}{Speedup} & \shortstack{Combine\\overlap} & \shortstack{Dispatch\\overlap} & \shortstack{Latency\\(ms/step)} & \multicolumn{1}{r!{\tabdashvline[15pt]}}{Speedup} & \shortstack{Combine\\overlap} & \shortstack{Dispatch\\overlap} \\
\midrule
\multirow{4}{*}{EP=4}
& Sequential & $291.9 \pm 0.2$ & 1.000$\times$ & 0\% & 0\% & $11.13 \pm 0.01$ & 1.000$\times$ & 0\% & 0\% \\
& Parallel & $262.5 \pm 0.5$ & 1.112$\times$ & 0\% & 93.2\% & $10.42 \pm 0.01$ & 1.068$\times$ & 0\% & 86.9\% \\
& Ladder & $267.8 \pm 0.2$ & 1.090$\times$ & 86.1\% & 11.6\% & $10.09 \pm 0.01$ & 1.103$\times$ & 92.8\% & 11.1\% \\
& BRAID & $\mathbf{249.2} \pm 0.4$ & \textbf{1.171$\times$} & 96.3\% & 92.8\% & $\mathbf{9.41} \pm 0.01$ & \textbf{1.182$\times$} & 89.2\% & 82.7\% \\
\midrule
\multirow{4}{*}{EP=8}
& Sequential & $479 \pm 3$ & 1.000$\times$ & 0\% & 0\% & $13.45 \pm 0.03$ & 1.000$\times$ & 0\% & 0\% \\
& Parallel & $408 \pm 3$ & 1.175$\times$ & 0\% & 52.9\% & $11.01 \pm 0.03$ & 1.222$\times$ & 0\% & 90.1\% \\
& Ladder & $423 \pm 3$ & 1.133$\times$ & 90.6\% & 2.6\% & $11.55 \pm 0.04$ & 1.165$\times$ & 65.2\% & 4.3\% \\
& BRAID & $\mathbf{407} \pm 2$ & \textbf{1.178$\times$} & 95.3\% & 31.3\% & $\mathbf{10.53} \pm 0.03$ & \textbf{1.278$\times$} & 88.8\% & 5.8\% \\
\bottomrule
\end{tabular}
\end{table*}

\begin{figure*}[t]
\centering
\pgfplotsset{every axis/.append style={width=4.35cm,height=4.0cm,xmode=log,
  log basis x=2,ymin=1.0,ymax=1.25,ytick={1.0,1.05,1.10,1.15,1.20,1.25},
  yticklabel style={/pgf/number format/fixed,/pgf/number format/precision=2},
  grid=major,grid style={black!10},tick label style={font=\scriptsize},
  label style={font=\scriptsize},title style={font=\small},
  every axis plot/.append style={line width=.9pt,mark size=1.6pt}}}
\begin{NoHyper}\ref{sweeplegend}\end{NoHyper}\\[2pt]
\begin{tikzpicture}
\begin{axis}[title={(a) Prefill, $L=1024$},xlabel={Batch $B$},xtick={16,64,256},xticklabels={16,64,256},ylabel={Speedup}]
\addplot[color=violet,mark=square*] coordinates {(16,1.094) (64,1.112) (256,1.096)};
\addplot[color=teal,mark=triangle*] coordinates {(16,1.052) (64,1.090) (256,1.088)};
\addplot[color=orange,mark=*] coordinates {(16,1.141) (64,1.171) (256,1.170)};
\end{axis}
\end{tikzpicture}\hfill
\begin{tikzpicture}
\begin{axis}[title={(b) Prefill, $B=64$},xlabel={Context $L$},xtick={256,1024,4096},xticklabels={256,1024,4096}]
\addplot[color=violet,mark=square*] coordinates {(256,1.081) (1024,1.112) (4096,1.105)};
\addplot[color=teal,mark=triangle*] coordinates {(256,1.045) (1024,1.090) (4096,1.078)};
\addplot[color=orange,mark=*] coordinates {(256,1.122) (1024,1.171) (4096,1.161)};
\end{axis}
\end{tikzpicture}\hfill
\begin{tikzpicture}
\begin{axis}[title={(c) Decode, $L=1024$},xlabel={Batch $B$},xtick={16,64,256},xticklabels={16,64,256}]
\addplot[color=violet,mark=square*] coordinates {(16,1.088) (64,1.068) (256,1.059)};
\addplot[color=teal,mark=triangle*] coordinates {(16,1.113) (64,1.103) (256,1.092)};
\addplot[color=orange,mark=*] coordinates {(16,1.223) (64,1.182) (256,1.136)};
\end{axis}
\end{tikzpicture}\hfill
\begin{tikzpicture}
\begin{axis}[title={(d) Decode, $B=64$},xlabel={Context $L$},xtick={256,1024,4096},xticklabels={256,1024,4096},legend to name=sweeplegend,legend columns=3,legend style={font=\scriptsize,draw=none,/tikz/every even column/.append style={column sep=8pt}}]
\addplot[color=violet,mark=square*] coordinates {(256,1.079) (1024,1.068) (4096,1.061)};
\addplot[color=teal,mark=triangle*] coordinates {(256,1.115) (1024,1.103) (4096,1.092)};
\addplot[color=orange,mark=*] coordinates {(256,1.203) (1024,1.182) (4096,1.160)};
\legend{Parallel,Ladder,BRAID}
\end{axis}
\end{tikzpicture}
\caption{Speedup relative to Sequential for EP=4 (one four-H100 node). (a) Prefill versus global batch size $B$ at $L=1024$; (b) prefill versus prompt length $L$ at $B=64$; (c) decode versus $B$ at $L=1024$; and (d) decode versus $L$ at $B=64$. Each curve represents one execution schedule; the shared legend applies to all four panels. Detailed latency and overlap measurements are reported in Appendix~\ref{app:scaling} (Table~\ref{tab:sweep}).}
\label{fig:sweeps}
\end{figure*}
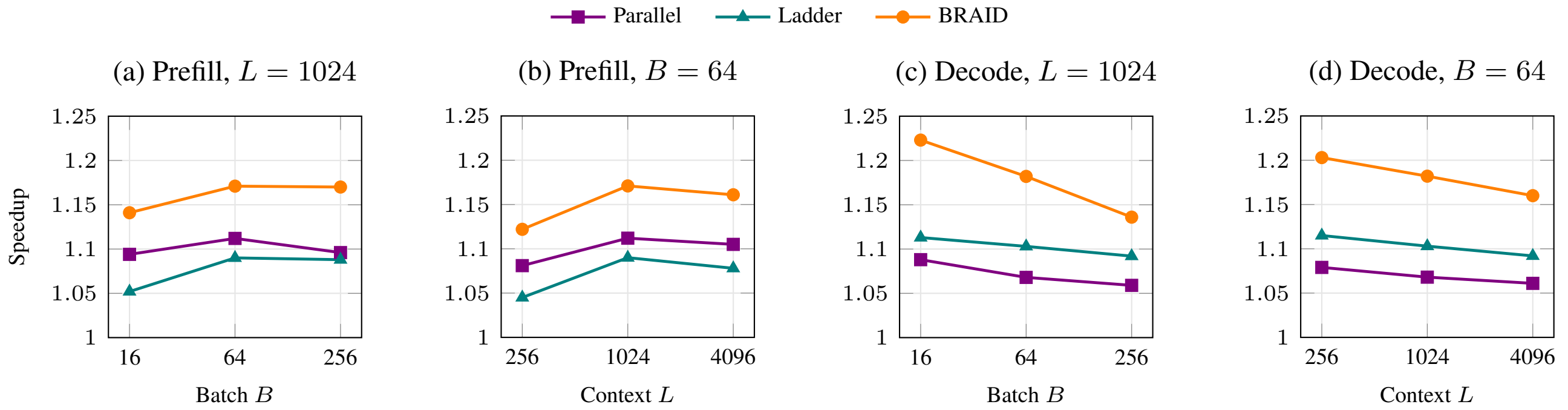

\begin{figure*}[t]
\centering
\pgfplotsset{every axis/.append style={width=4.35cm,height=4.0cm,xmode=log,
  log basis x=2,ymin=1.0,ymax=1.40,ytick={1.0,1.10,1.20,1.30,1.40},
  yticklabel style={/pgf/number format/fixed,/pgf/number format/precision=2},
  grid=major,grid style={black!10},tick label style={font=\scriptsize},
  label style={font=\scriptsize},title style={font=\small},
  every axis plot/.append style={line width=.9pt,mark size=1.6pt}}}
\begin{NoHyper}\ref{sweeplegendeight}\end{NoHyper}\\[2pt]
\begin{tikzpicture}
\begin{axis}[title={(a) Prefill, $L=1024$},xlabel={Batch $B$},xmax=600,xtick={32,128,512},xticklabels={32,128,512},ylabel={Speedup}]
\addplot[color=violet,mark=square*] coordinates {(32,1.143) (128,1.175) (512,1.190)};
\addplot[color=teal,mark=triangle*] coordinates {(32,1.113) (128,1.133) (512,1.135)};
\addplot[color=orange,mark=*] coordinates {(32,1.150) (128,1.178) (512,1.200)};
\end{axis}
\end{tikzpicture}\hfill
\begin{tikzpicture}
\begin{axis}[title={(b) Prefill, $B=128$},xlabel={Context $L$},xtick={256,1024,4096},xticklabels={256,1024,4096}]
\addplot[color=violet,mark=square*] coordinates {(256,1.128) (1024,1.175) (4096,1.271)};
\addplot[color=teal,mark=triangle*] coordinates {(256,1.107) (1024,1.133) (4096,1.120)};
\addplot[color=orange,mark=*] coordinates {(256,1.136) (1024,1.178) (4096,1.287)};
\end{axis}
\end{tikzpicture}\hfill
\begin{tikzpicture}
\begin{axis}[title={(c) Decode, $L=1024$},xlabel={Batch $B$},xmax=600,xtick={32,128,512},xticklabels={32,128,512}]
\addplot[color=violet,mark=square*] coordinates {(32,1.208) (128,1.222) (512,1.209)};
\addplot[color=teal,mark=triangle*] coordinates {(32,1.118) (128,1.165) (512,1.244)};
\addplot[color=orange,mark=*] coordinates {(32,1.219) (128,1.278) (512,1.365)};
\end{axis}
\end{tikzpicture}\hfill
\begin{tikzpicture}
\begin{axis}[title={(d) Decode, $B=128$},xlabel={Context $L$},xtick={256,1024,4096},xticklabels={256,1024,4096},legend to name=sweeplegendeight,legend columns=3,legend style={font=\scriptsize,draw=none,/tikz/every even column/.append style={column sep=8pt}}]
\addplot[color=violet,mark=square*] coordinates {(256,1.201) (1024,1.222) (4096,1.202)};
\addplot[color=teal,mark=triangle*] coordinates {(256,1.146) (1024,1.165) (4096,1.226)};
\addplot[color=orange,mark=*] coordinates {(256,1.253) (1024,1.278) (4096,1.370)};
\legend{Parallel,Ladder,BRAID}
\end{axis}
\end{tikzpicture}
\caption{Speedup relative to Sequential for EP=8 (two four-H100 nodes). (a) Prefill versus global batch size $B$ at $L=1024$; (b) prefill versus prompt length $L$ at $B=128$; (c) decode versus $B$ at $L=1024$; and (d) decode versus $L$ at $B=128$. Each curve represents one execution schedule; the shared legend applies to all four panels. Detailed latency and overlap measurements are reported in Appendix~\ref{app:scaling} (Table~\ref{tab:sweep8}).}
\label{fig:sweeps8}
\end{figure*}

\paragraph{Prefill.}
During prefill, communication scales approximately linearly with the number of tokens $BL$, while attention computation scales linearly with $B$ and quadratically with $L$. Increasing context length $L$ therefore increases the computation available to overlap with communication. For EP=4, BRAID's speedup rises from $1.12\times$ at $L=256$ to approximately $1.16$--$1.17\times$ for $L\geq1024$, as attention accounts for an increasing share of execution time and provides a longer computation window to overlap intra-node communication (Figure~\ref{fig:sweeps}(b)). Varying batch size $B$ at fixed $L$ has a weaker effect, since attention, expert computation, and communication all scale approximately linearly with $B$, leaving their relative costs, and hence BRAID's speedup over Sequential, largely unchanged (Figure~\ref{fig:sweeps}(a)). For EP=8, inter-node communication remains substantially longer than the available computation window, limiting the fraction that can be hidden. As $L$ increases, the longer attention window allows Parallel and BRAID to overlap more of the dispatch, while Ladder's combine is already mostly overlapped at $L=1024$, causing its speedup to saturate (Figure~\ref{fig:sweeps8}(b)). At $L=4096$, BRAID achieves $1.29\times$, compared with $1.27\times$ for Parallel and $1.12\times$ for Ladder. As for EP=4, increasing $B$ has a weaker effect on speedup (Figure~\ref{fig:sweeps8}(a)).

\paragraph{Decode.}
Decode processes one new token per request, so communication volume depends only on $B$. At the tested workloads, communication is largely latency-bound, dominated by fixed overheads rather than data transfer: its duration varies little with $L$ and grows with $B$ mainly at the largest EP=8 batch (Table~\ref{tab:timing}), whereas decode attention scales approximately with $BL$. For EP=4, the computation window already exceeds the short intra-node communication, so speedups decrease as $B$ or $L$ grows (Figures~\ref{fig:sweeps}(c)--(d)). BRAID nevertheless leads by overlapping both collectives. For EP=8, the longer inter-node communication exceeds the available computation window. Increasing $B$ or $L$ allows BRAID to hide more communication and Ladder to hide more of its combine, whereas Parallel's dispatch is already mostly overlapped, keeping its speedup near $1.21\times$. BRAID's speedup increases from $1.22\times$ to approximately $1.37\times$ (Figures~\ref{fig:sweeps8}(c)--(d)), reaching a substantial advantage over Parallel and Ladder at the largest $B$ and $L$ tested.

Full latency, phase-timing, and overlap results are given in Appendix~\ref{app:inference} (Tables~\ref{tab:timing}, \ref{tab:sweep}, and~\ref{tab:sweep8}).

\section{Conclusion}
BRAID-MoE combines early routing with laddered residuals to relax the sequential combine--route--dispatch dependency. By computing expert assignments from the pre-combine residual while preserving fresh activations for expert computation, BRAID exposes new opportunities for overlapping expert-parallel communication with computation. Our experiments show that this additional parallelism comes at a limited quality cost. At matched pretraining budgets, BRAID's validation loss remains within 0.98\% and 0.49\% of the sequential baseline for 0.4B-A0.1B and 2.5B-A0.5B, respectively. Converting the final half of the Transformer layers in pretrained sequential models to BRAID with only 10\% additional training yields validation losses within 0.9\% and 0.3\% of the original sequential checkpoints at 0.4B-A0.1B and 2.5B-A0.5B, respectively. On an instruction-tuned OLMoE-1B-7B checkpoint, supervised fine-tuning on only 5\% of the Infinity-Instruct 7M and Gen subsets recovers average downstream accuracy from 60.2\% after conversion to 70.2\%, compared with 74.6\% for the original checkpoint. In controlled inference experiments on Qwen3-30B-A3B, BRAID outperforms all competing schedules, speeding up prefill and decode by $1.17\times$ and $1.18\times$ on our reference EP=4 workload, and by up to $1.29\times$ and $1.37\times$ across two nodes (EP=8).

These results demonstrate that limited routing staleness can be exploited to improve communication--computation overlap while largely preserving model quality. The resulting adjacency between combine and dispatch creates opportunities for kernel fusion and further overlap with tensor-parallel collectives. Exploring these optimizations remains a direction for future work.

\section*{Acknowledgements}
JZ, DA and EO were supported by the PEPR IA SHARP grant (ANR-23-PEIA-0008) and benefited from access to the Jean Zay HPC/AI resources of IDRIS under allocations AD011015884R1, AD011016641R1, and AD011017766 made by GENCI. EB acknowledges funding from NSERC Discovery.

\bibliography{references}

\begin{thebibliography}{46}
\providecommand{\natexlab}[1]{#1}
\providecommand{\url}[1]{\texttt{#1}}
\expandafter\ifx\csname urlstyle\endcsname\relax
  \providecommand{\doi}[1]{doi: #1}\else
  \providecommand{\doi}{doi: \begingroup \urlstyle{rm}\Url}\fi

\bibitem[Ahmed \& Oyallon(2026)Ahmed and Oyallon]{ahmed2026know}
Ahmed, D. and Oyallon, E.
\newblock Know what you need: Efficient activation checkpointing.
\newblock \emph{The Fortieth Annual Conference on Neural Information Processing Systems}, 2026.

\bibitem[Ansel et~al.(2024)Ansel, Yang, He, Gimelshein, Jain, Voznesensky, Bao, Bell, Berard, Burovski, Chauhan, Chourdia, Constable, Desmaison, DeVito, Ellison, Feng, Gong, Gschwind, Hirsh, Huang, Kalambarkar, Kirsch, Lazos, Lezcano, Liang, Liang, Lu, Luk, Maher, Pan, Puhrsch, Reso, Saroufim, Siraichi, Suk, Zhang, Suo, Tillet, Zhao, Wang, Zhou, Zou, Wang, Mathews, Wen, Chanan, Wu, and Chintala]{ansel2024pytorch}
Ansel, J., Yang, E., He, H., Gimelshein, N., Jain, A., Voznesensky, M., Bao, B., Bell, P., Berard, D., Burovski, E., Chauhan, G., Chourdia, A., Constable, W., Desmaison, A., DeVito, Z., Ellison, E., Feng, W., Gong, J., Gschwind, M., Hirsh, B., Huang, S., Kalambarkar, K., Kirsch, L., Lazos, M., Lezcano, M., Liang, Y., Liang, J., Lu, Y., Luk, C.~K., Maher, B., Pan, Y., Puhrsch, C., Reso, M., Saroufim, M., Siraichi, M.~Y., Suk, H., Zhang, S., Suo, M., Tillet, P., Zhao, X., Wang, E., Zhou, K., Zou, R., Wang, X., Mathews, A., Wen, W., Chanan, G., Wu, P., and Chintala, S.
\newblock {{PyTorch} 2: Faster Machine Learning Through Dynamic {Python} Bytecode Transformation and Graph Compilation}.
\newblock In \emph{Proceedings of the 29th ACM International Conference on Architectural Support for Programming Languages and Operating Systems, Volume 2}, ASPLOS '24, pp.\  929--947. ACM, April 2024.
\newblock \doi{10.1145/3620665.3640366}.
\newblock URL \url{https://doi.org/10.1145/3620665.3640366}.

\bibitem[Belilovsky et~al.(2026)Belilovsky, Horoi, Oyallon, and Th{\'e}rien]{belilovsky2026inner}
Belilovsky, E., Horoi, S., Oyallon, E., and Th{\'e}rien, B.
\newblock How inner optimizers shape replica scaling.
\newblock \emph{arXiv preprint arXiv:2610.19064}, 2026.

\bibitem[Black et~al.(2022)Black, Biderman, Hallahan, Anthony, Gao, Golding, He, Leahy, McDonell, Phang, Pieler, Prashanth, Purohit, Reynolds, Tow, Wang, and Weinbach]{black-etal-2022-gpt}
Black, S., Biderman, S., Hallahan, E., Anthony, Q., Gao, L., Golding, L., He, H., Leahy, C., McDonell, K., Phang, J., Pieler, M., Prashanth, U.~S., Purohit, S., Reynolds, L., Tow, J., Wang, B., and Weinbach, S.
\newblock {GPT}-{N}eo{X}-20{B}: An open-source autoregressive language model.
\newblock In Fan, A., Ilic, S., Wolf, T., and Gall{\'e}, M. (eds.), \emph{Proceedings of {B}ig{S}cience Episode {\#}5 -- Workshop on Challenges {\&} Perspectives in Creating Large Language Models}, pp.\  95--136, virtual+Dublin, May 2022. Association for Computational Linguistics.
\newblock \doi{10.18653/v1/2022.bigscience-1.9}.
\newblock URL \url{https://aclanthology.org/2022.bigscience-1.9/}.

\bibitem[Cagnasso et~al.(2026{\natexlab{a}})Cagnasso, Belilovsky, and Oyallon]{cagnasso2026unifyinglocalcommunicationslocal}
Cagnasso, P., Belilovsky, E., and Oyallon, E.
\newblock Unifying local communications and local updates for llm pretraining, 2026{\natexlab{a}}.
\newblock URL \url{https://arxiv.org/abs/2606.11081}.

\bibitem[Cagnasso et~al.(2026{\natexlab{b}})Cagnasso, Kopyshevskiy, Oyallon, and Belilovsky]{cagnasso2026spicesparseparametergossip}
Cagnasso, P., Kopyshevskiy, D., Oyallon, E., and Belilovsky, E.
\newblock Spice: Sparse parameter gossip for llm pretraining, 2026{\natexlab{b}}.
\newblock URL \url{https://arxiv.org/abs/2611.23053}.

\bibitem[Cai et~al.(2025)Cai, Jiang, Qin, Cui, Kim, and Huang]{cai2025shortcut}
Cai, W., Jiang, J., Qin, L., Cui, J., Kim, S., and Huang, J.
\newblock Shortcut-connected expert parallelism for accelerating mixture of experts.
\newblock In Singh, A., Fazel, M., Hsu, D., Lacoste-Julien, S., Berkenkamp, F., Maharaj, T., Wagstaff, K., and Zhu, J. (eds.), \emph{Proceedings of the 42nd International Conference on Machine Learning}, volume 267 of \emph{Proceedings of Machine Learning Research}, pp.\  6211--6228. PMLR, 13--19 Jul 2025.
\newblock URL \url{https://proceedings.mlr.press/v267/cai25c.html}.

\bibitem[Chang et~al.(2024)Chang, Bao, Hou, Jiang, Zheng, Zhong, Zhang, Song, Yao, Jiang, Lin, Jin, and Liu]{chang2024flux}
Chang, L.-W., Bao, W., Hou, Q., Jiang, C., Zheng, N., Zhong, Y., Zhang, X., Song, Z., Yao, C., Jiang, Z., Lin, H., Jin, X., and Liu, X.
\newblock Flux: Fast software-based communication overlap on gpus through kernel fusion, 2024.
\newblock URL \url{https://arxiv.org/abs/2406.06858}.

\bibitem[Chowdhery et~al.(2023)Chowdhery, Narang, Devlin, Bosma, Mishra, Roberts, Barham, Chung, Sutton, Gehrmann, Schuh, Shi, Tsvyashchenko, Maynez, Rao, Barnes, Tay, Shazeer, Prabhakaran, Reif, Du, Hutchinson, Pope, Bradbury, Austin, Isard, Gur-Ari, Yin, Duke, Levskaya, Ghemawat, Dev, Michalewski, Garcia, Misra, Robinson, Fedus, Zhou, Ippolito, Luan, Lim, Zoph, Spiridonov, Sepassi, Dohan, Agrawal, Omernick, Dai, Pillai, Pellat, Lewkowycz, Moreira, Child, Polozov, Lee, Zhou, Wang, Saeta, Diaz, Firat, Catasta, Wei, Meier-Hellstern, Eck, Dean, Petrov, and Fiedel]{chowdhery2023palm}
Chowdhery, A., Narang, S., Devlin, J., Bosma, M., Mishra, G., Roberts, A., Barham, P., Chung, H.~W., Sutton, C., Gehrmann, S., Schuh, P., Shi, K., Tsvyashchenko, S., Maynez, J., Rao, A., Barnes, P., Tay, Y., Shazeer, N., Prabhakaran, V., Reif, E., Du, N., Hutchinson, B., Pope, R., Bradbury, J., Austin, J., Isard, M., Gur-Ari, G., Yin, P., Duke, T., Levskaya, A., Ghemawat, S., Dev, S., Michalewski, H., Garcia, X., Misra, V., Robinson, K., Fedus, L., Zhou, D., Ippolito, D., Luan, D., Lim, H., Zoph, B., Spiridonov, A., Sepassi, R., Dohan, D., Agrawal, S., Omernick, M., Dai, A.~M., Pillai, T.~S., Pellat, M., Lewkowycz, A., Moreira, E., Child, R., Polozov, O., Lee, K., Zhou, Z., Wang, X., Saeta, B., Diaz, M., Firat, O., Catasta, M., Wei, J., Meier-Hellstern, K., Eck, D., Dean, J., Petrov, S., and Fiedel, N.
\newblock Palm: Scaling language modeling with pathways.
\newblock \emph{Journal of Machine Learning Research}, 24\penalty0 (240):\penalty0 1--113, 2023.
\newblock URL \url{http://jmlr.org/papers/v24/22-1144.html}.

\bibitem[Cui et~al.(2026)Cui, Jackson, Herrera, T\'{a}rano, and Kerner]{cui2026multihead}
Cui, C., Jackson, R., Herrera, B.~J., T\'{a}rano, A.~M., and Kerner, H.
\newblock Multi-head {L}atent{M}o{E} and head parallel: Communication-efficient and deterministic {M}o{E} parallelism.
\newblock In Zhang, T., Dudik, M., Jaggi, M., Agarwal, A., Li, S., Schuurmans, D., Zhu, J., Berkenkamp, F., Dong, H., and Bietti, A. (eds.), \emph{Proceedings of the 43rd International Conference on Machine Learning}, volume 306 of \emph{Proceedings of Machine Learning Research}, pp.\  21847--21863. PMLR, 06--11 Jul 2026.
\newblock URL \url{https://proceedings.mlr.press/v306/cui26c.html}.

\bibitem[DeepSeek-AI et~al.(2025)DeepSeek-AI, Liu, Feng, Xue, Wang, Wu, Lu, Zhao, Deng, Zhang, Ruan, Dai, Guo, Yang, Chen, Ji, Li, Lin, Dai, Luo, Hao, Chen, Li, Zhang, Bao, Xu, Wang, Zhang, Ding, Xin, Gao, Li, Qu, Cai, Liang, Guo, Ni, Li, Wang, Chen, Chen, Yuan, Qiu, Li, Song, Dong, Hu, Gao, Guan, Huang, Yu, Wang, Zhang, Xu, Xia, Zhao, Wang, Zhang, Li, Wang, Zhang, Zhang, Tang, Li, Tian, Huang, Wang, Zhang, Wang, Zhu, Chen, Du, Chen, Jin, Ge, Zhang, Pan, Wang, Xu, Zhang, Chen, Li, Lu, Zhou, Chen, Wu, Ye, Ye, Ma, Wang, Zhou, Yu, Zhou, Pan, Wang, Yun, Pei, Sun, Xiao, Zeng, Zhao, An, Liu, Liang, Gao, Yu, Zhang, Li, Jin, Wang, Bi, Liu, Wang, Shen, Chen, Zhang, Chen, Nie, Sun, Wang, Cheng, Liu, Xie, Liu, Yu, Song, Shan, Zhou, Yang, Li, Su, Lin, Li, Wang, Wei, Zhu, Zhang, Xu, Xu, Huang, Li, Zhao, Sun, Li, Wang, Yu, Zheng, Zhang, Shi, Xiong, He, Tang, Piao, Wang, Tan, Ma, Liu, Guo, Wu, Ou, Zhu, Wang, Gong, Zou, He, Zha, Xiong, Ma, Yan, Luo, You, Liu, Zhou, Wu, Ren, Ren, Sha, Fu, Xu, Huang, Zhang, Xie, Zhang, Hao,
  Gou, Ma, Yan, Shao, Xu, Wu, Zhang, Li, Gu, Zhu, Liu, Li, Xie, Song, Gao, and Pan]{deepseekai2025deepseekv3technicalreport}
DeepSeek-AI, Liu, A., Feng, B., Xue, B., Wang, B., Wu, B., Lu, C., Zhao, C., Deng, C., Zhang, C., Ruan, C., Dai, D., Guo, D., Yang, D., Chen, D., Ji, D., Li, E., Lin, F., Dai, F., Luo, F., Hao, G., Chen, G., Li, G., Zhang, H., Bao, H., Xu, H., Wang, H., Zhang, H., Ding, H., Xin, H., Gao, H., Li, H., Qu, H., Cai, J.~L., Liang, J., Guo, J., Ni, J., Li, J., Wang, J., Chen, J., Chen, J., Yuan, J., Qiu, J., Li, J., Song, J., Dong, K., Hu, K., Gao, K., Guan, K., Huang, K., Yu, K., Wang, L., Zhang, L., Xu, L., Xia, L., Zhao, L., Wang, L., Zhang, L., Li, M., Wang, M., Zhang, M., Zhang, M., Tang, M., Li, M., Tian, N., Huang, P., Wang, P., Zhang, P., Wang, Q., Zhu, Q., Chen, Q., Du, Q., Chen, R.~J., Jin, R.~L., Ge, R., Zhang, R., Pan, R., Wang, R., Xu, R., Zhang, R., Chen, R., Li, S.~S., Lu, S., Zhou, S., Chen, S., Wu, S., Ye, S., Ye, S., Ma, S., Wang, S., Zhou, S., Yu, S., Zhou, S., Pan, S., Wang, T., Yun, T., Pei, T., Sun, T., Xiao, W.~L., Zeng, W., Zhao, W., An, W., Liu, W., Liang, W., Gao, W., Yu, W., Zhang, W.,
  Li, X.~Q., Jin, X., Wang, X., Bi, X., Liu, X., Wang, X., Shen, X., Chen, X., Zhang, X., Chen, X., Nie, X., Sun, X., Wang, X., Cheng, X., Liu, X., Xie, X., Liu, X., Yu, X., Song, X., Shan, X., Zhou, X., Yang, X., Li, X., Su, X., Lin, X., Li, Y.~K., Wang, Y.~Q., Wei, Y.~X., Zhu, Y.~X., Zhang, Y., Xu, Y., Xu, Y., Huang, Y., Li, Y., Zhao, Y., Sun, Y., Li, Y., Wang, Y., Yu, Y., Zheng, Y., Zhang, Y., Shi, Y., Xiong, Y., He, Y., Tang, Y., Piao, Y., Wang, Y., Tan, Y., Ma, Y., Liu, Y., Guo, Y., Wu, Y., Ou, Y., Zhu, Y., Wang, Y., Gong, Y., Zou, Y., He, Y., Zha, Y., Xiong, Y., Ma, Y., Yan, Y., Luo, Y., You, Y., Liu, Y., Zhou, Y., Wu, Z.~F., Ren, Z.~Z., Ren, Z., Sha, Z., Fu, Z., Xu, Z., Huang, Z., Zhang, Z., Xie, Z., Zhang, Z., Hao, Z., Gou, Z., Ma, Z., Yan, Z., Shao, Z., Xu, Z., Wu, Z., Zhang, Z., Li, Z., Gu, Z., Zhu, Z., Liu, Z., Li, Z., Xie, Z., Song, Z., Gao, Z., and Pan, Z.
\newblock Deepseek-v3 technical report, 2025.
\newblock URL \url{https://arxiv.org/abs/2412.19437}.

\bibitem[Dukler et~al.(2026)Dukler, Li, Shah, Liu, Appia, and Barsoum]{dukler2026farskip}
Dukler, Y., Li, G., Shah, D., Liu, J., Appia, V., and Barsoum, E.
\newblock Farskip-collective: Unhobbling blocking communication in mixture of experts models.
\newblock In Chowdhery, A. and Jia, Z. (eds.), \emph{Proceedings of Machine Learning and Systems}, volume~8, pp.\  1803--1816. MLSys, 2026.
\newblock URL \url{https://proceedings.mlsys.org/paper_files/paper/2026/file/6feb9b30798abcfae937760d183605e1-Paper-Conference.pdf}.

\bibitem[Elango et~al.(2026)Elango, Bhatia, Waleffe, Shafipour, Asida, Khattar, Assaf, Golub, Guman, Mitra, Zhao, Borkar, Zilberstein, Patwary, Shoeybi, and Darvish~Rouhani]{elango2026latentmoe}
Elango, V., Bhatia, N., Waleffe, R., Shafipour, R., Asida, T., Khattar, A., Assaf, N., Golub, M., Guman, J., Mitra, T., Zhao, R., Borkar, R., Zilberstein, R., Patwary, M., Shoeybi, M., and Darvish~Rouhani, B.
\newblock Revisiting efficiency–accuracy scaling in mixture-of-experts architectures.
\newblock In Zhang, T., Dudik, M., Jaggi, M., Agarwal, A., Li, S., Schuurmans, D., Zhu, J., Berkenkamp, F., Dong, H., and Bietti, A. (eds.), \emph{Proceedings of the 43rd International Conference on Machine Learning}, volume 306 of \emph{Proceedings of Machine Learning Research}, pp.\  27809--27821. PMLR, 06--11 Jul 2026.
\newblock URL \url{https://proceedings.mlr.press/v306/elango26a.html}.

\bibitem[Fedus et~al.(2022)Fedus, Zoph, and Shazeer]{fedus2022switch}
Fedus, W., Zoph, B., and Shazeer, N.
\newblock Switch transformers: Scaling to trillion parameter models with simple and efficient sparsity.
\newblock \emph{Journal of Machine Learning Research}, 23\penalty0 (120):\penalty0 1--39, 2022.
\newblock URL \url{http://jmlr.org/papers/v23/21-0998.html}.

\bibitem[Gu et~al.(2025)Gu, Tafjord, Kuehl, Haddad, Dodge, and Hajishirzi]{gu2025olmes}
Gu, Y., Tafjord, O., Kuehl, B., Haddad, D., Dodge, J., and Hajishirzi, H.
\newblock {OLMES}: A standard for language model evaluations.
\newblock In Chiruzzo, L., Ritter, A., and Wang, L. (eds.), \emph{Findings of the {A}ssociation for {C}omputational {L}inguistics: {NAACL} 2025}, pp.\  5020--5048, Albuquerque, New Mexico, April 2025. Association for Computational Linguistics.
\newblock ISBN 979-8-89176-195-7.
\newblock \doi{10.18653/v1/2025.findings-naacl.282}.
\newblock URL \url{https://aclanthology.org/2025.findings-naacl.282/}.

\bibitem[Guo et~al.(2026)Guo, Mishra, Cheng, Stoica, and Dao]{guo2026sonicmoe}
Guo, W., Mishra, M., Cheng, X., Stoica, I., and Dao, T.
\newblock Sonicmoe: Accelerating moe with io and tile-aware optimizations.
\newblock In Vondrick, C., Hariharan, B., Raffel, C., Pinto, L., Yang, D., and Faust, A. (eds.), \emph{International Conference on Learning Representations}, volume 2026, pp.\  67814--67839, 2026.
\newblock URL \url{https://proceedings.iclr.cc/paper_files/paper/2026/file/6ed5bf446f59e2c6646d23058c86424b-Paper-Conference.pdf}.

\bibitem[Hwang et~al.(2023)Hwang, Cui, Xiong, Yang, Liu, Hu, Wang, Salas, Jose, Ram, Chau, Cheng, Yang, Yang, and Xiong]{hwang2023tutel}
Hwang, C., Cui, W., Xiong, Y., Yang, Z., Liu, Z., Hu, H., Wang, Z., Salas, R., Jose, J., Ram, P., Chau, H., Cheng, P., Yang, F., Yang, M., and Xiong, Y.
\newblock Tutel: Adaptive mixture-of-experts at scale.
\newblock In Song, D., Carbin, M., and Chen, T. (eds.), \emph{Proceedings of Machine Learning and Systems}, volume~5, pp.\  269--287. Curan, 2023.
\newblock URL \url{https://proceedings.mlsys.org/paper_files/paper/2023/file/5616d34cf8ff73942cfd5aa922842556-Paper-mlsys2023.pdf}.

\bibitem[Hwang et~al.(2024)Hwang, Wei, Cao, Hwang, Tang, Cao, and Yang]{hwang2024pregated}
Hwang, R., Wei, J., Cao, S., Hwang, C., Tang, X., Cao, T., and Yang, M.
\newblock Pre-gated moe: An algorithm-system co-design for fast and scalable mixture-of-expert inference.
\newblock In \emph{2024 ACM/IEEE 51st Annual International Symposium on Computer Architecture (ISCA)}, pp.\  1018--1031, 2024.
\newblock \doi{10.1109/ISCA59077.2024.00078}.

\bibitem[Janson et~al.(2026{\natexlab{a}})Janson, Oyallon, and Belilovsky]{janson2026learned}
Janson, P., Oyallon, E., and Belilovsky, E.
\newblock Learned subspace compression for communication-efficient pipeline parallelism.
\newblock \emph{arXiv preprint arXiv:2606.05484}, 2026{\natexlab{a}}.

\bibitem[Janson et~al.(2026{\natexlab{b}})Janson, Oyallon, and Belilovsky]{janson2026stabilizing}
Janson, P., Oyallon, E., and Belilovsky, E.
\newblock Stabilizing native low-rank {LLM} pretraining.
\newblock In \emph{Proceedings of the 43rd International Conference on Machine Learning}, volume 306 of \emph{Proceedings of Machine Learning Research}, pp.\  50830--50852. PMLR, 06--11 Jul 2026{\natexlab{b}}.
\newblock URL \url{https://proceedings.mlr.press/v306/janson26a.html}.

\bibitem[Jiang et~al.(2024)Jiang, Tian, Jia, Zheng, Wu, and Wang]{jiang2024lancet}
Jiang, C., Tian, Y., Jia, Z., Zheng, S., Wu, C., and Wang, Y.
\newblock Lancet: Accelerating mixture-of-experts training via whole graph computation-communication overlapping.
\newblock In Gibbons, P., Pekhimenko, G., and Sa, C.~D. (eds.), \emph{Proceedings of Machine Learning and Systems}, volume~6, pp.\  74--86, 2024.
\newblock URL \url{https://proceedings.mlsys.org/paper_files/paper/2024/file/339caf45a6fa281cae8adc6465343464-Paper-Conference.pdf}.

\bibitem[Jordan et~al.(2024)Jordan, Jin, Boza, You, Cesista, Newhouse, and Bernstein]{jordan2024muon}
Jordan, K., Jin, Y., Boza, V., You, J., Cesista, F., Newhouse, L., and Bernstein, J.
\newblock Muon: An optimizer for hidden layers in neural networks, 2024.
\newblock URL \url{https://kellerjordan.github.io/posts/muon/}.

\bibitem[Kopyshevskiy et~al.(2026)Kopyshevskiy, Th{\'e}rien, Pinto, and Oyallon]{kopyshevskiy2026}
Kopyshevskiy, D., Th{\'e}rien, B., Pinto, N., and Oyallon, E.
\newblock {LACCO: Local Accumulation with Computation--Communication Overlap}.
\newblock working paper or preprint, November 2026.
\newblock URL \url{https://hal.science/hal-05862888}.

\bibitem[Kwon et~al.(2023)Kwon, Li, Zhuang, Sheng, Zheng, Yu, Gonzalez, Zhang, and Stoica]{kwon2023efficient}
Kwon, W., Li, Z., Zhuang, S., Sheng, Y., Zheng, L., Yu, C.~H., Gonzalez, J., Zhang, H., and Stoica, I.
\newblock Efficient memory management for large language model serving with pagedattention.
\newblock In \emph{Proceedings of the 29th Symposium on Operating Systems Principles}, SOSP '23, pp.\  611–626, New York, NY, USA, 2023. Association for Computing Machinery.
\newblock ISBN 9798400702297.
\newblock \doi{10.1145/3600006.3613165}.
\newblock URL \url{https://doi.org/10.1145/3600006.3613165}.

\bibitem[Lepikhin et~al.(2021)Lepikhin, Lee, Xu, Chen, Firat, Huang, Krikun, Shazeer, and Chen]{lepikhin2021gshard}
Lepikhin, D., Lee, H., Xu, Y., Chen, D., Firat, O., Huang, Y., Krikun, M., Shazeer, N., and Chen, Z.
\newblock {\{}GS{\}}hard: Scaling giant models with conditional computation and automatic sharding.
\newblock In \emph{International Conference on Learning Representations}, 2021.
\newblock URL \url{https://openreview.net/forum?id=qrwe7XHTmYb}.

\bibitem[Li et~al.(2023)Li, Jiang, Zhu, Wang, and Xu]{li2023lina}
Li, J., Jiang, Y., Zhu, Y., Wang, C., and Xu, H.
\newblock Accelerating distributed {MoE} training and inference with lina.
\newblock In \emph{2023 USENIX Annual Technical Conference (USENIX ATC 23)}, pp.\  945--959, Boston, MA, July 2023. USENIX Association.
\newblock ISBN 978-1-939133-35-9.
\newblock URL \url{https://www.usenix.org/conference/atc23/presentation/li-jiamin}.

\bibitem[Liang et~al.(2025)Liang, Liu, Wright, Constable, Gu, Huang, Zhang, Feng, Huang, Wang, Purandare, Nadathur, and Idreos]{liang2025torchtitan}
Liang, W., Liu, T., Wright, L., Constable, W., Gu, A., Huang, C.-C., Zhang, I., Feng, W., Huang, H., Wang, J., Purandare, S.~J., Nadathur, G., and Idreos, S.
\newblock Torchtitan: One-stop pytorch native solution for production ready llm pretraining.
\newblock In Yue, Y., Garg, A., Peng, N., Sha, F., and Yu, R. (eds.), \emph{International Conference on Learning Representations}, volume 2025, pp.\  92161--92181, 2025.
\newblock URL \url{https://proceedings.iclr.cc/paper_files/paper/2025/file/e6231c5f46598cfd09ff1970524e0436-Paper-Conference.pdf}.

\bibitem[Liu et~al.(2025)Liu, Su, Yao, Jiang, Lai, Du, Qin, Xu, Lu, Yan, Chen, Zheng, Liu, Liu, Yin, He, Zhu, Wang, Wang, Dong, Zhang, Kang, Zhang, Xu, Zhang, Wu, Zhou, and Yang]{liu2025muonscalable}
Liu, J., Su, J., Yao, X., Jiang, Z., Lai, G., Du, Y., Qin, Y., Xu, W., Lu, E., Yan, J., Chen, Y., Zheng, H., Liu, Y., Liu, S., Yin, B., He, W., Zhu, H., Wang, Y., Wang, J., Dong, M., Zhang, Z., Kang, Y., Zhang, H., Xu, X., Zhang, Y., Wu, Y., Zhou, X., and Yang, Z.
\newblock Muon is scalable for llm training, 2025.
\newblock URL \url{https://arxiv.org/abs/2502.16982}.

\bibitem[Loshchilov \& Hutter(2017)Loshchilov and Hutter]{loshchilov2017sgdr}
Loshchilov, I. and Hutter, F.
\newblock {SGDR}: Stochastic gradient descent with warm restarts.
\newblock In \emph{International Conference on Learning Representations}, 2017.
\newblock URL \url{https://openreview.net/forum?id=Skq89Scxx}.

\bibitem[Loshchilov \& Hutter(2019)Loshchilov and Hutter]{loshchilov2019decoupled}
Loshchilov, I. and Hutter, F.
\newblock Decoupled weight decay regularization.
\newblock In \emph{International Conference on Learning Representations}, 2019.
\newblock URL \url{https://openreview.net/forum?id=Bkg6RiCqY7}.

\bibitem[Ludziejewski et~al.(2025)Ludziejewski, Pi\'{o}ro, Krajewski, Stefaniak, Krutul, Ma{\l}a\'{s}nicki, Cygan, Sankowski, Adamczewski, Mi{\l}o\'{s}, and Jaszczur]{ludziejewski2025joint}
Ludziejewski, J., Pi\'{o}ro, M., Krajewski, J., Stefaniak, M., Krutul, M., Ma{\l}a\'{s}nicki, J., Cygan, M., Sankowski, P., Adamczewski, K., Mi{\l}o\'{s}, P., and Jaszczur, S.
\newblock Joint {M}o{E} scaling laws: Mixture of experts can be memory efficient.
\newblock In Singh, A., Fazel, M., Hsu, D., Lacoste-Julien, S., Berkenkamp, F., Maharaj, T., Wagstaff, K., and Zhu, J. (eds.), \emph{Proceedings of the 42nd International Conference on Machine Learning}, volume 267 of \emph{Proceedings of Machine Learning Research}, pp.\  41056--41073. PMLR, 13--19 Jul 2025.
\newblock URL \url{https://proceedings.mlr.press/v267/ludziejewski25a.html}.

\bibitem[Muennighoff et~al.(2025)Muennighoff, Soldaini, Groeneveld, Lo, Morrison, Min, Shi, Walsh, Tafjord, Lambert, Gu, Arora, Bhagia, Schwenk, Wadden, Wettig, Hui, Dettmers, Kiela, Farhadi, Smith, Koh, Singh, and Hajishirzi]{muennighoff2025olmoe}
Muennighoff, N., Soldaini, L., Groeneveld, D., Lo, K., Morrison, J., Min, S., Shi, W., Walsh, P., Tafjord, O., Lambert, N., Gu, Y., Arora, S., Bhagia, A., Schwenk, D., Wadden, D., Wettig, A., Hui, B., Dettmers, T., Kiela, D., Farhadi, A., Smith, N., Koh, P.~W., Singh, A., and Hajishirzi, H.
\newblock Olmoe: Open mixture-of-experts language models.
\newblock In Yue, Y., Garg, A., Peng, N., Sha, F., and Yu, R. (eds.), \emph{International Conference on Learning Representations}, volume 2025, pp.\  62061--62121, 2025.
\newblock URL \url{https://proceedings.iclr.cc/paper_files/paper/2025/file/9b224ace8963c9385ad5e2b5c9039b97-Paper-Conference.pdf}.

\bibitem[Nabli et~al.(2025)Nabli, Fournier, ERBACHER, Serrano, Belilovsky, and Oyallon]{nabli2025acco}
Nabli, A., Fournier, L., ERBACHER, P., Serrano, L., Belilovsky, E., and Oyallon, E.
\newblock Acco: Accumulate while you communicate for communication-overlapped sharded llm training.
\newblock In Belgrave, D., Zhang, C., Lin, H., Pascanu, R., Koniusz, P., Ghassemi, M., and Chen, N. (eds.), \emph{Advances in Neural Information Processing Systems}, volume 38, Main Conference, pp.\  69278--69305. Curran Associates, Inc., 2025.
\newblock \doi{10.52202/085713-2329}.
\newblock URL \url{https://proceedings.neurips.cc/paper_files/paper/2025/file/6454dcd80b5373daaa97e53ce32c78a1-Paper-Conference.pdf}.

\bibitem[Penedo et~al.(2024)Penedo, Kydl\'{\i}\v{c}ek, allal, Lozhkov, Mitchell, Raffel, Von~Werra, and Wolf]{penedo2024fineweb}
Penedo, G., Kydl\'{\i}\v{c}ek, H., allal, L.~B., Lozhkov, A., Mitchell, M., Raffel, C., Von~Werra, L., and Wolf, T.
\newblock The fineweb datasets: Decanting the web for the finest text data at scale.
\newblock In Globerson, A., Mackey, L., Belgrave, D., Fan, A., Paquet, U., Tomczak, J., and Zhang, C. (eds.), \emph{Advances in Neural Information Processing Systems}, volume~37, pp.\  30811--30849. Curran Associates, Inc., 2024.
\newblock \doi{10.52202/079017-0970}.
\newblock URL \url{https://proceedings.neurips.cc/paper_files/paper/2024/file/370df50ccfdf8bde18f8f9c2d9151bda-Paper-Datasets_and_Benchmarks_Track.pdf}.

\bibitem[Qiu et~al.(2025)Qiu, Huang, Cheng, Zhou, wang, Titov, and Fu]{qiu2025rmoe}
Qiu, Z., Huang, Z., Cheng, S., Zhou, Y., wang, z., Titov, I., and Fu, J.
\newblock Layerwise recurrent router for mixture-of-experts.
\newblock In Yue, Y., Garg, A., Peng, N., Sha, F., and Yu, R. (eds.), \emph{International Conference on Learning Representations}, volume 2025, pp.\  14179--14205, 2025.
\newblock URL \url{https://proceedings.iclr.cc/paper_files/paper/2025/file/250a6c6a7a7216f5dec1364e9a93abf4-Paper-Conference.pdf}.

\bibitem[Qiu et~al.(2026)Qiu, Wang, Li, Li, Xu, Wang, Zhang, Men, Mao, Zhang, Zhou, Luo, Huang, Lian, Huang, Chen, Zhang, Xu, Wang, Chen, Wang, Jiang, Yuan, Sun, Jin, Zhang, Wang, Ren, Wang, Zhang, Dong, Cao, Ma, Mao, Zheng, and Liu]{qwen2026design}
Qiu, Z., Wang, Z., Li, X., Li, Y., Xu, Y., Wang, Y., Zhang, H., Men, R., Mao, B., Zhang, C., Zhou, F., Luo, H., Huang, H., Lian, H., Huang, H., Chen, H., Zhang, J., Xu, J., Wang, J., Chen, L., Wang, L., Jiang, L., Yuan, M., Sun, M., Jin, P., Zhang, S., Wang, S., Ren, X., Wang, Y., Zhang, Y., Dong, Y., Cao, Y., Ma, Y., Mao, Y., Zheng, B., and Liu, D.
\newblock On the design of {Qwen3.8-Next} architecture: Evaluation, efficiency, and training stability, 2026.
\newblock URL \url{https://arxiv.org/abs/2608.30320}.

\bibitem[Rivaud et~al.(2025)Rivaud, Fournier, Pumir, Belilovsky, Eickenberg, and Oyallon]{rivaud2025petra}
Rivaud, S., Fournier, L., Pumir, T., Belilovsky, E., Eickenberg, M., and Oyallon, E.
\newblock Petra: Parallel end-to-end training with reversible architectures.
\newblock In Yue, Y., Garg, A., Peng, N., Sha, F., and Yu, R. (eds.), \emph{International Conference on Learning Representations}, volume 2025, pp.\  80162--80178, 2025.
\newblock URL \url{https://proceedings.iclr.cc/paper_files/paper/2025/file/c710d6b4507e70c6332bee871b8d1ca5-Paper-Conference.pdf}.

\bibitem[Shazeer et~al.(2017)Shazeer, Mirhoseini, Maziarz, Davis, Le, Hinton, and Dean]{shazeer2017outrageously}
Shazeer, N., Mirhoseini, A., Maziarz, K., Davis, A., Le, Q., Hinton, G., and Dean, J.
\newblock Outrageously large neural networks: The sparsely-gated mixture-of-experts layer.
\newblock In \emph{International Conference on Learning Representations}, 2017.
\newblock URL \url{https://openreview.net/forum?id=B1ckMDqlg}.

\bibitem[Su et~al.(2024)Su, Ahmed, Lu, Pan, Bo, and Liu]{su2024roformer}
Su, J., Ahmed, M., Lu, Y., Pan, S., Bo, W., and Liu, Y.
\newblock Roformer: Enhanced transformer with rotary position embedding.
\newblock \emph{Neurocomputing}, 568:\penalty0 127063, 2024.
\newblock ISSN 0925-2312.
\newblock \doi{https://doi.org/10.1016/j.neucom.2023.127063}.
\newblock URL \url{https://www.sciencedirect.com/science/article/pii/S0925231223011864}.

\bibitem[Team et~al.(2026)Team, Bai, Bai, Bao, C., Cai, Cai, Cao, Cao, Chai, Charles, Che, Chen, Chen, Chen, Chen, Chen, Chen, Chen, Chen, Chen, Chen, Chen, Chen, Chen, Chen, Chen, Chen, Chen, Chen, Chen, Chen, Cheng, Cheng, Cui, Cui, Dai, Deng, Ding, Ding, Ding, Dong, Dong, Dong, Dong, Du, Du, Du, Du, Du, Fan, Feng, Feng, Feng, Fu, Fu, Gao, Gao, Gao, Gao, Gao, Geng, Gong, Gong, Gong, Gong, Gu, Gu, Guan, Guo, Guo, Guo, Guo, Hao, Hao, Hao, He, He, He, He, He, He, He, He, He, Hong, Hong, Hu, Hu, Hu, Hu, Hu, Hu, Hua, Huang, Huang, Huang, Huang, Huang, Huang, Huang, Huang, Hui, Jia, Jiang, Jiang, Jiang, Jin, Jin, Jing, Kong, Lai, Li, Li, Li, Li, Li, Li, Li, Li, Li, Li, Li, Li, Li, Li, Li, Li, Li, Li, Li, Li, Li, Li, Li, Li, Lin, Lin, Lin, Lin, Lin, Liu, Liu, Liu, Liu, Liu, Liu, Liu, Liu, Liu, Liu, Liu, Liu, Liu, Liu, Liu, Lu, Lu, Lu, Lu, Lu, Luo, Luo, Luo, Luo, Lyu, Lyu, Mao, Mei, Men, Ni, Niu, Pan, Peng, Qi, Qin, Qin, Qin, Qiu, Qiu, Qiu, Qu, Qu, Shang, Shao, Shen, Shi, Shi, Shi, Shi, Siu, Song, Song, Su, Su, Su,
  Sui, Sun, Sun, Sun, Sun, Sun, Sun, Tai, Tang, Tang, Tang, Tang, Tian, Tian, Tian, Tu, Wang, Wang, Wang, Wang, Wang, Wang, Wang, Wang, Wang, Wang, Wang, Wang, Wang, Wang, Wang, Wang, Wang, Wang, Wang, Wang, Wang, Wang, Wang, Wang, Wang, Wang, Wang, Wang, Wang, Wang, Wang, Wang, Wang, Wang, Wang, Wang, Wang, Wang, Wang, Wang, Wang, Wei, Wei, Wei, Wen, Wu, Wu, Wu, Wu, Wu, Wu, Wu, Wu, Wu, Xian, Xiang, Xiang, Xiao, Xiao, Xiao, Xie, Xie, Xie, Xie, Xing, Xiong, Xu, Xu, Xu, Xu, Xu, Xu, Xu, Xu, Xu, Xu, Xu, Xu, Xu, Xu, Xu, Xu, Xu, Xu, Xue, Yan, Yan, Yang, Yang, Yang, Yang, Yang, Yang, Yang, Yang, Yang, Yang, Yang, Yang, Yang, Yang, Yang, Yang, Yao, Ye, Ye, Ye, Ye, Yin, Yin, Yin, Yu, Yu, Yu, Yu, Yu, Yu, Yuan, Yuan, Yue, Yue, Yue, Zha, Zhan, Zhang, Zhang, Zhang, Zhang, Zhang, Zhang, Zhang, Zhang, Zhang, Zhang, Zhang, Zhang, Zhang, Zhang, Zhang, Zhang, Zhang, Zhang, Zhang, Zhang, Zhang, Zhang, Zhang, Zhang, Zhang, Zhang, Zhang, Zhang, Zhang, Zhang, Zhang, Zhao, Zhao, Zhao, Zhao, Zhao, Zhao, Zhao, Zhao, Zhao, Zhao, Zhao,
  Zheng, Zheng, Zheng, Zheng, Zheng, Zhong, Zhong, Zhong, Zhou, Zhou, Zhou, Zhou, Zhou, Zhou, Zhou, Zhu, Zhu, Zhu, Zhu, Zhu, Zhu, Zhuang, Zhuang, and Zu]{kimi2026k3}
Team, K., Bai, T., Bai, Y., Bao, Y., C., M., Cai, J., Cai, X., Cao, P., Cao, Y., Chai, Z., Charles, Y., Che, H.~S., Chen, G., Chen, G., Chen, G., Chen, H., Chen, J., Chen, J., Chen, J., Chen, K., Chen, P., Chen, R., Chen, W., Chen, X., Chen, Y., Chen, Y., Chen, Y., Chen, Y., Chen, Y., Chen, Y., Chen, Y., Chen, Z., Cheng, D., Cheng, Y., Cui, J., Cui, J., Dai, A., Deng, J., Ding, H., Ding, R., Ding, S., Dong, M., Dong, M., Dong, Y., Dong, Y., Du, A., Du, C., Du, D., Du, J., Du, Y., Fan, Y., Feng, J., Feng, Q., Feng, Y., Fu, K., Fu, Q., Gao, F., Gao, H., Gao, J., Gao, T., Gao, W., Geng, S., Gong, J., Gong, L., Gong, S., Gong, X., Gu, Q., Gu, Y., Guan, S., Guo, H., Guo, S., Guo, X., Guo, Z., Hao, B., Hao, W., Hao, X., He, D., He, H., He, L., He, Q., He, W., He, X., He, X., He, Y., He, Y., Hong, C., Hong, T., Hu, H., Hu, J., Hu, R., Hu, W., Hu, Y., Hu, Z., Hua, L., Huang, J., Huang, K., Huang, R., Huang, S., Huang, W., Huang, Y., Huang, Z., Huang, Z., Hui, Y., Jia, C., Jiang, Y., Jiang, Z., Jiang, Z., Jin, W.,
  Jin, X., Jing, Y., Kong, H., Lai, G., Li, A., Li, C., Li, C., Li, C., Li, F., Li, G., Li, H., Li, J., Li, J., Li, L., Li, L., Li, L., Li, W., Li, W., Li, X., Li, Y., Li, Y., Li, Y., Li, Y., Li, Z., Li, Z., Li, Z., Li, Z., Li, Z., Lin, J., Lin, X., Lin, Y., Lin, Z., Lin, Z., Liu, B., Liu, B., Liu, C., Liu, L., Liu, S., Liu, S., Liu, S., Liu, T., Liu, W., Liu, Y., Liu, Y., Liu, Y., Liu, Y., Liu, Z., Liu, Z., Lu, E., Lu, H., Lu, L., Lu, T., Lu, Z., Luo, A., Luo, G., Luo, J., Luo, Y., Lyu, B., Lyu, W., Mao, S., Mei, Y., Men, X., Ni, M., Niu, Y., Pan, S., Peng, S., Qi, Z., Qin, R., Qin, Z., Qin, Z., Qiu, H., Qiu, J., Qiu, J., Qu, B., Qu, Y., Shang, Z., Shao, Y., Shen, H., Shi, J., Shi, J., Shi, L., Shi, S., Siu, W., Song, P., Song, X., Su, J., Su, Y., Su, Z., Sui, L., Sun, J., Sun, J., Sun, S., Sun, S., Sun, T., Sun, Y., Tai, Y., Tang, C., Tang, H., Tang, S., Tang, Z., Tian, C., Tian, R., Tian, Y., Tu, W., Wang, C., Wang, C., Wang, C., Wang, D., Wang, F., Wang, H., Wang, H., Wang, H., Wang, H., Wang, H., Wang,
  H., Wang, J., Wang, J., Wang, J., Wang, J., Wang, L., Wang, S., Wang, S., Wang, S., Wang, S., Wang, S., Wang, T., Wang, W., Wang, X., Wang, X., Wang, X., Wang, X., Wang, Y., Wang, Y., Wang, Y., Wang, Y., Wang, Y., Wang, Y., Wang, Y., Wang, Y., Wang, Z., Wang, Z., Wang, Z., Wang, Z., Wang, Z., Wang, Z., Wei, C., Wei, M., Wei, S., Wen, Z., Wu, F., Wu, H., Wu, R., Wu, W., Wu, X., Wu, Y., Wu, Y., Wu, Y., Wu, Z., Xian, X., Xiang, C., Xiang, Y., Xiao, B., Xiao, C., Xiao, X., Xie, J., Xie, X., Xie, Y., Xie, Z., Xing, B., Xiong, Y., Xu, B., Xu, B., Xu, J., Xu, J., Xu, J., Xu, J., Xu, L.~H., Xu, Q., Xu, S., Xu, S., Xu, T., Xu, T., Xu, W., Xu, X., Xu, Y., Xu, Y., Xu, Y., Xu, Z., Xue, H., Yan, J., Yan, Y., Yang, F., Yang, G., Yang, H., Yang, J., Yang, R., Yang, W., Yang, X., Yang, X., Yang, Y., Yang, Y., Yang, Y., Yang, Y., Yang, Z., Yang, Z., Yang, Z., Yang, Z., Yao, H., Ye, D., Ye, H., Ye, W., Ye, Z., Yin, B., Yin, H., Yin, X., Yu, C., Yu, H., Yu, L., Yu, S., Yu, S., Yu, T., Yuan, E., Yuan, M., Yue, T., Yue, W.,
  Yue, Y., Zha, D., Zhan, H., Zhang, B.~H., Zhang, D., Zhang, F., Zhang, H., Zhang, H., Zhang, H., Zhang, J., Zhang, J., Zhang, J., Zhang, K., Zhang, M., Zhang, P., Zhang, Q., Zhang, R., Zhang, R., Zhang, S., Zhang, S., Zhang, X., Zhang, X., Zhang, Y., Zhang, Y., Zhang, Y., Zhang, Y., Zhang, Y., Zhang, Y., Zhang, Y., Zhang, Y., Zhang, Y., Zhang, Y., Zhang, Z., Zhang, Z., Zhao, B., Zhao, C., Zhao, F., Zhao, J., Zhao, J., Zhao, S., Zhao, W., Zhao, X., Zhao, X., Zhao, Y., Zhao, Z., Zheng, H., Zheng, H., Zheng, R., Zheng, S., Zheng, T., Zhong, H., Zhong, L., Zhong, L., Zhou, M., Zhou, Q., Zhou, R., Zhou, R., Zhou, X., Zhou, Y., Zhou, Z., Zhu, J., Zhu, L., Zhu, X., Zhu, Y., Zhu, Y., Zhu, Z., Zhuang, C., Zhuang, W., and Zu, X.
\newblock Kimi k3: Open frontier intelligence, 2026.
\newblock URL \url{https://arxiv.org/abs/2607.24653}.

\bibitem[Touvron et~al.(2023)Touvron, Martin, Stone, Albert, Almahairi, Babaei, Bashlykov, Batra, Bhargava, Bhosale, Bikel, Blecher, Ferrer, Chen, Cucurull, Esiobu, Fernandes, Fu, Fu, Fuller, Gao, Goswami, Goyal, Hartshorn, Hosseini, Hou, Inan, Kardas, Kerkez, Khabsa, Kloumann, Korenev, Koura, Lachaux, Lavril, Lee, Liskovich, Lu, Mao, Martinet, Mihaylov, Mishra, Molybog, Nie, Poulton, Reizenstein, Rungta, Saladi, Schelten, Silva, Smith, Subramanian, Tan, Tang, Taylor, Williams, Kuan, Xu, Yan, Zarov, Zhang, Fan, Kambadur, Narang, Rodriguez, Stojnic, Edunov, and Scialom]{touvron2023llama2}
Touvron, H., Martin, L., Stone, K., Albert, P., Almahairi, A., Babaei, Y., Bashlykov, N., Batra, S., Bhargava, P., Bhosale, S., Bikel, D., Blecher, L., Ferrer, C.~C., Chen, M., Cucurull, G., Esiobu, D., Fernandes, J., Fu, J., Fu, W., Fuller, B., Gao, C., Goswami, V., Goyal, N., Hartshorn, A., Hosseini, S., Hou, R., Inan, H., Kardas, M., Kerkez, V., Khabsa, M., Kloumann, I., Korenev, A., Koura, P.~S., Lachaux, M.-A., Lavril, T., Lee, J., Liskovich, D., Lu, Y., Mao, Y., Martinet, X., Mihaylov, T., Mishra, P., Molybog, I., Nie, Y., Poulton, A., Reizenstein, J., Rungta, R., Saladi, K., Schelten, A., Silva, R., Smith, E.~M., Subramanian, R., Tan, X.~E., Tang, B., Taylor, R., Williams, A., Kuan, J.~X., Xu, P., Yan, Z., Zarov, I., Zhang, Y., Fan, A., Kambadur, M., Narang, S., Rodriguez, A., Stojnic, R., Edunov, S., and Scialom, T.
\newblock Llama 2: Open foundation and fine-tuned chat models, 2023.
\newblock URL \url{https://arxiv.org/abs/2307.09288}.

\bibitem[{vLLM Project}(2026)]{vllm_dbo}
{vLLM Project}.
\newblock {Dual Batch Overlap}.
\newblock \url{https://docs.vllm.ai/en/latest/design/dbo/}, 2026.
\newblock vLLM Documentation. Accessed: 2026-10-30.

\bibitem[Yang et~al.(2025)Yang, Li, Yang, Zhang, Hui, Zheng, Yu, Gao, Huang, Lv, Zheng, Liu, Zhou, Huang, Hu, Ge, Wei, Lin, Tang, Yang, Tu, Zhang, Yang, Yang, Zhou, Zhou, Lin, Dang, Bao, Yang, Yu, Deng, Li, Xue, Li, Zhang, Wang, Zhu, Men, Gao, Liu, Luo, Li, Tang, Yin, Ren, Wang, Zhang, Ren, Fan, Su, Zhang, Zhang, Wan, Liu, Wang, Cui, Zhang, Zhou, and Qiu]{qwen2025qwen3}
Yang, A., Li, A., Yang, B., Zhang, B., Hui, B., Zheng, B., Yu, B., Gao, C., Huang, C., Lv, C., Zheng, C., Liu, D., Zhou, F., Huang, F., Hu, F., Ge, H., Wei, H., Lin, H., Tang, J., Yang, J., Tu, J., Zhang, J., Yang, J., Yang, J., Zhou, J., Zhou, J., Lin, J., Dang, K., Bao, K., Yang, K., Yu, L., Deng, L., Li, M., Xue, M., Li, M., Zhang, P., Wang, P., Zhu, Q., Men, R., Gao, R., Liu, S., Luo, S., Li, T., Tang, T., Yin, W., Ren, X., Wang, X., Zhang, X., Ren, X., Fan, Y., Su, Y., Zhang, Y., Zhang, Y., Wan, Y., Liu, Y., Wang, Z., Cui, Z., Zhang, Z., Zhou, Z., and Qiu, Z.
\newblock Qwen3 technical report, 2025.
\newblock URL \url{https://arxiv.org/abs/2505.09388}.

\bibitem[Zhang et~al.(2025{\natexlab{a}})Zhang, Mishra, Zhou, Brandon, Wang, Kim, Ragan-Kelley, Song, Athiwaratkun, and Dao]{zhang2025ladder}
Zhang, M., Mishra, M., Zhou, Z., Brandon, W., Wang, J., Kim, Y., Ragan-Kelley, J., Song, S.~L., Athiwaratkun, B., and Dao, T.
\newblock Ladder-residual: Parallelism-aware architecture for accelerating large model inference with communication overlapping.
\newblock In Singh, A., Fazel, M., Hsu, D., Lacoste-Julien, S., Berkenkamp, F., Maharaj, T., Wagstaff, K., and Zhu, J. (eds.), \emph{Proceedings of the 42nd International Conference on Machine Learning}, volume 267 of \emph{Proceedings of Machine Learning Research}, pp.\  75779--75792. PMLR, 13--19 Jul 2025{\natexlab{a}}.
\newblock URL \url{https://proceedings.mlr.press/v267/zhang25bg.html}.

\bibitem[Zhang et~al.(2025{\natexlab{b}})Zhang, Zheng, Lin, Jiang, Bao, Jiang, Hou, Cui, Zheng, Chang, Chen, and Liu]{zhang2025comet}
Zhang, S., Zheng, N., Lin, H., Jiang, Z., Bao, W., Jiang, C., Hou, Q., Cui, W., Zheng, S., Chang, L.-W., Chen, Q., and Liu, X.
\newblock Comet: Fine-grained computation-communication overlapping for mixture-of-experts.
\newblock In Zaharia, M., Joshi, G., and Lin, Y. (eds.), \emph{Proceedings of Machine Learning and Systems}, volume~7. MLSys, 2025{\natexlab{b}}.
\newblock URL \url{https://proceedings.mlsys.org/paper_files/paper/2025/file/e27ea0cd50b798ff8942caf9203f0992-Paper-Conference.pdf}.

\bibitem[Zhao et~al.(2025)Zhao, Zhou, Zhang, Deng, Xu, Liu, Yu, Li, and Zhao]{deepseek2025deepep}
Zhao, C., Zhou, S., Zhang, L., Deng, C., Xu, Z., Liu, Y., Yu, K., Li, J., and Zhao, L.
\newblock Deepep: an efficient expert-parallel communication library.
\newblock \url{https://github.com/deepseek-ai/DeepEP}, 2025.

\end{thebibliography}
\bibliographystyle{mlsys2025}

\section*{AI Usage Disclosure}
ChatGPT was used to assist with literature review, manuscript drafting and editing. Codex and Claude Code were used for code generation and review.

\newpage
\appendix

\section{Experimental Hardware}
\label{app:hardware}

All experiments use the same cluster of four-GPU H100 SXM5 nodes (80\,GB HBM3 and 132 SMs per GPU). Each GPU has 18 NVLink~4 links, six to each other GPU in the node, providing up to 300\,GB/s bidirectional bandwidth per GPU pair (150\,GB/s per direction), or 900\,GB/s aggregate bidirectional bandwidth per GPU. Inter-node communication uses four ConnectX-7 NDR400 InfiniBand adapters per node, each rated at 50\,GB/s per direction (200\,GB/s per direction per node). Pretraining and checkpoint continuation use one four-GPU node; post-training recovery uses two GPUs; and inference uses one node for EP=4 or two nodes for EP=8.

\section{Architecture and Training Details}
\label{app:details}

\subsection{Architectures and Initialization}

Our 0.4B-A0.1B and 2.5B-A0.5B Qwen3-style models preserve selected proportions of Qwen3-MoE~\citep{qwen2025qwen3}. We keep head dimension $d_h=128$, total query width $h_qd_h=2d$, and expert intermediate width $m=3d/8$. The depth and model width are chosen to reach the target active-parameter scales. Both models use 64 routed experts, top-8 routing, and no shared experts. Table~\ref{tab:models} summarizes their geometry.

\begin{table}[t]
\centering\small
\caption{Qwen3-shaped model configurations used for the controlled experiments.}
\label{tab:models}
\setlength{\tabcolsep}{3pt}
\begin{tabular*}{\linewidth}{@{\extracolsep{\fill}}lrrrrr@{}}
\toprule
Scale & Layers & $d$ & \shortstack{Q/KV\\heads}
& \shortstack{Expert\\width} & Experts \\
\midrule
0.4B-A0.1B & 14 & 576 & 9/1 & 216 & 64 \\
2.5B-A0.5B & 24 & 1152 & 18/2 & 432 & 64 \\
\bottomrule
\end{tabular*}
\end{table}

The backbone uses pre-RMSNorm with $\epsilon=10^{-6}$, Q/K RMS normalization, SwiGLU experts, and rotary position embeddings (RoPE)~\citep{su2024roformer} with base $10^6$. The input embedding and language-model head are untied, with a vocabulary of 32,000 tokens. Routing uses a softmax over all experts, followed by top-8 selection and renormalization of the selected probabilities to sum to one. In BRAID, the early router has a separate RMSNorm from the normalization used for the fresh expert input.

We follow the TorchTitan Qwen3-style initialization~\citep{liang2025torchtitan}. Embedding weights, the language-model head, attention input projections, router weights, and expert input/up projections are initialized from $\mathcal{N}(0,0.02^2)$. Attention output projections and expert down projections use standard deviation $0.02/\sqrt{2N_{\mathrm{layers}}}$, where $N_{\mathrm{layers}}$ is the number of Transformer layers. All normalization weights, including the separate early-router normalization, are initialized to one.

\subsection{Model--Token Allocation and Data Streams}
\label{app:model-token-allocation}

For the pretraining experiments in Section~\ref{sec:pretraining}, we use the Joint MoE scaling-law prescription~\citep{ludziejewski2025joint} to determine the training budgets. Although the original fit assumes a different routing configuration, a local IsoFLOP sweep at the smaller scale finds that the 100M model performs best among the tested 75M, 100M, and 125M configurations. We therefore apply the same prescription at 2.5B-A0.5B, treating it as a reproducible budgeting rule rather than evidence of global compute optimality. The resulting training horizons are 3,123 and 9,974 optimization steps, respectively.

\paragraph{Data streams.}
We draw pretraining and validation tokens from the CC-MAIN-2022-05 Common Crawl snapshot of FineWeb v1.4.0~\citep{penedo2024fineweb}, which deduplicates documents within each snapshot. Training and held-out validation use disjoint parquet files to prevent overlap between the two sets.

\subsection{Pretraining Optimization}
Our pretraining setup broadly follows \citet{ahmed2026know}, with adjustments for MoE training. Attention and expert matrices use Muon~\citep{jordan2024muon}, with momentum 0.95, Nesterov momentum, and five Newton--Schulz iterations. We replace Muon's original update scaling with update-RMS scaling~\citep{liu2025muonscalable}. Embeddings, the language-model head, routers, and normalization parameters use AdamW~\citep{loshchilov2019decoupled} with $(\beta_1,\beta_2)=(0.9,0.95)$. Following \citet{qwen2026design}, who report that applying Muon to MoE routers exacerbates early-training fluctuations without providing significant performance gains, we retain AdamW for the router parameters. Normalization parameters are excluded from weight decay, and gradients are clipped at global norm 1.0. Load balancing uses an expert-bias update with coefficient $10^{-3}$. No differentiable load-balancing loss is used, and tokens are not dropped.

Learning rates use 5\% linear warmup followed by cosine decay~\citep{loshchilov2017sgdr} to 10\% of the peak rate. We select the peak learning rate for each scale using the sequential reference, searching over a $\{1,5,7\}\times10^j$ grid. The selected learning rate and weight decay are then held fixed across architectures at the same scale. At 0.4B-A0.1B, the Muon and AdamW peak learning rates are both $7\times10^{-3}$, with 156 warmup steps over the 3,123-step schedule. At 2.5B-A0.5B, both peak learning rates are $5\times10^{-3}$, with 498 warmup steps over 9,974 steps. Weight decay is 0.1 at both scales. All training runs use BF16.

\subsection{Checkpoint Continuation}
\label{app:checkpoint-continuation}

For the checkpoint adaptation experiments in Section~\ref{sec:checkpoint-adaptation}, at 0.4B-A0.1B BRAID uses a peak learning rate of $2.1\times10^{-3}$ ($0.3\times$ the pretraining peak) and the sequential control $7\times10^{-4}$ ($0.1\times$), both with 15 warmup steps. At 2.5B-A0.5B, BRAID uses a peak learning rate of $1.5\times10^{-3}$ ($0.3\times$ the pretraining peak)  and the sequential control $5\times10^{-4}$ ($0.1\times$), both with 49 warmup steps. Both continuations use weight decay 0.1 and cosine decay to 10\% of the peak learning rate. Continuation data are drawn from separate parquet files of the CC-MAIN-2022-05 snapshot of FineWeb v1.4.0, disjoint from pretraining and held-out validation.

\subsection{OLMoE Supervised Fine-tuning}
\label{app:olmoe-sft}

For the post-training recovery experiment in Section~\ref{sec:post-training-recovery}, the sampled corpus contains 444,913 conversations, preserving the original subset proportions and multi-turn conversations. Of the approximately 205.8M tokens per epoch, 138.3M assistant tokens contribute to the training loss. We use a maximum sequence length of 4,096, zero weight decay, and gradient clipping at global norm 1.0. Training uses BF16 and DeepSpeed ZeRO-3.

\section{Inference Implementation and Profiling Details}
\label{app:inference}

This appendix documents the execution, workload, and profiling choices behind Section~\ref{sec:perf}, followed by a breakdown of the observed speedups.

\subsection{Implementation and Experimental Controls}
\label{app:inference-details}

\paragraph{Execution engine.} The four schedules are implemented as a custom vLLM~\citep{kwon2023efficient} model class whose forward pass runs our two-stream schedule code. vLLM schedules requests asynchronously, manages the paged key-value (KV) cache, runs attention with its FlashAttention backend, and samples greedily. Attention uses data parallelism and experts use EP: each GPU runs one engine with tensor parallelism disabled and holds $1/\mathrm{EP}$ of the requests and $128/\mathrm{EP}$ experts per layer. Dispatch and combine use DeepEP's high-throughput kernels with one buffer sized for the whole local prefill. Chunked prefill and prefix caching are disabled, so each GPU prefills its local batch in a single step; the KV cache is sized explicitly so that no request is preempted, and end-of-sequence tokens are ignored so that every request generates exactly 32 tokens.

\paragraph{Compilation and CUDA graphs.}
We disable vLLM's graph compiler because it splits the forward pass at attention and cannot trace cross-layer stream switches and DeepEP calls. Instead, element-wise and normalization regions (routing, fused residual additions and RMSNorms, query/key normalization with RoPE) are compiled separately with \texttt{torch.compile}~\citep{ansel2024pytorch} (full graph, static shapes), reusing one compiled function across layers with weights as inputs. Experts run outside compilation as two grouped GEMMs around a Triton SwiGLU kernel. Prefill runs without CUDA graphs and uses CPU synchronization to size DeepEP receive buffers exactly. Decode avoids this synchronization by allocating buffers for the full local batch and capturing the entire forward pass, including both streams and DeepEP kernels, in one CUDA graph per step. Graphs are recaptured for each schedule and SM setting, with recording excluded from timing.

\paragraph{Weights and routing.}
The BF16 checkpoint is loaded once per GPU, including attention, normalization, router, embedding, and local expert weights, with gate and up projections concatenated. Routing follows Qwen3: FP32 softmax over 128 experts, top-8 selection, and renormalization. Since DeepEP's combine sums expert outputs without weighting, each received row is scaled by its FP32 routing weight inside the SwiGLU kernel, before the down projection. To isolate execution differences, all schedules replay expert indices and weights recorded by an untimed Sequential run on the same prompts. Sequential replay reproduces its original generated tokens.

\paragraph{Fusions.}\phantomsection\label{par:fusions}
Each schedule exploits the fusions its dependencies allow. Sequential's combine epilogue performs the expert reduction, residual addition, and next attention normalization, while its attention residual addition, MoE-input normalization, and routing form one compiled region. Parallel's combine epilogue adds both branches to the residual and computes both next-layer normalizations in a single dual-output pass. These normalizations share the residual input, as in parallel Transformer blocks~\citep{black-etal-2022-gpt}. Ladder and BRAID share a combine epilogue that performs the expert reduction and residual addition, producing both the updated residual and its normalized expert input. BRAID also fuses attention and early-router normalizations, computing the RMS statistic once and applying separate learned scales. In Ladder and BRAID, the attention residual addition precedes the wait on dispatch, as in Algorithm~\ref{alg:braid}.

\paragraph{Workloads.}
Prompts are fixed-length token windows drawn from a subset of FineWeb~v1.4.0, using the native Qwen3-30B-A3B-Base tokenizer. We vary batch size $B$ at fixed $L=1024$ and prompt length $L$ at fixed 16 prompts per GPU (Table~\ref{tab:workloads}). For EP=8, the global batch size is doubled relative to EP=4 to maintain the same number of prompts per GPU.

\begin{table}[t]
\centering\small
\caption{Inference workloads: prompt length $L$ and batch $B$ for EP=4 and EP=8.}
\label{tab:workloads}
\setlength{\tabcolsep}{5pt}
\begin{tabular}{rrrr}
\toprule
$L$ & $B$ (EP=4) & $B$ (EP=8) & Prompts per GPU \\
\midrule
1,024 & 16  & 32  & 4 \\
256   & 64  & 128 & 16 \\
1,024 & 64  & 128 & 16 \\
4,096 & 64  & 128 & 16 \\
1,024 & 256 & 512 & 64 \\
\bottomrule
\end{tabular}
\end{table}

\paragraph{Expert placement.} We compute the same static expert placement for all schedules using DeepSeek's EPLB~\citep{deepseekai2025deepseekv3technicalreport}, without expert replication. Expert loads are measured on 4.2M held-out FineWeb tokens; experts are assigned to GPUs in decreasing order of load, choosing the least-loaded GPU with an available slot. The placement is fixed during inference and permutes the expert weights and router rows at load time. Table~\ref{tab:imbalance} reports the resulting load balance.

\begin{table}[t]
\centering\small
\caption{Load of the busiest GPU relative to the mean on the benchmark prompts, averaged over layers (prefill).}
\label{tab:imbalance}
\setlength{\tabcolsep}{3.5pt}
\begin{tabular}{lrrrrr}
\toprule
$L$ & 1,024 & 256 & 1,024 & 4,096 & 1,024 \\
Prompts per GPU & 4 & 16 & 16 & 16 & 64 \\
\midrule
EP=4, contiguous & 1.208 & 1.214 & 1.202 & 1.197 & 1.200 \\
EP=4, EPLB         & 1.067 & 1.042 & 1.028 & 1.016 & 1.013 \\
EP=8, EPLB         & 1.086 & 1.059 & 1.032 & 1.025 & 1.024 \\
\bottomrule
\end{tabular}
\end{table}

\subsection{Communication Kernel Tuning}
\label{app:tuning}

\paragraph{SM allocation and resource contention.}\phantomsection\label{par:contention}
We illustrate the effect of resource contention on communication--computation overlap with a simple model. Suppose two kernels take times $t_1$ and $t_2$ in isolation, and their runtimes scale inversely with their allocated SM shares. Running them concurrently with fractions $\lambda$ and $1-\lambda$ gives
\begin{equation*}
\min_{0<\lambda<1}\max\!\left(\frac{t_1}{\lambda},\frac{t_2}{1-\lambda}\right)=t_1+t_2.
\end{equation*}
Under these assumptions, even optimal resource allocation offers no speedup over sequential execution. However, communication and computation primarily rely on different resources: interconnect bandwidth versus compute throughput or HBM bandwidth. This allows beneficial overlap despite some SM contention. Allocating more SMs to communication may reduce its latency while slowing concurrent computation. We therefore tune the SM allocations of DeepEP's transfer and epilogue phases to minimize end-to-end latency rather than communication latency alone.

\paragraph{SM tuning.}
For each schedule, we tune five parameters: the SM allocations of DeepEP's dispatch and combine transfers and their epilogues, and a compute SM margin $m$ that limits the SMs available to attention kernels. The search covers DeepEP's default setting, transfer allocations from 2 to 96 SMs, epilogues of 16, 32, 64, or all SMs, and margins from 0 to 64 SMs. Candidate configurations are evaluated on held-out prompts disjoint from the benchmark data, across all five workloads of each topology, using two warm-ups and seven randomized rounds. We select one configuration per schedule, topology, and phase (prefill or decode) that minimizes the worst relative slowdown against the fastest configuration on each workload. Table~\ref{tab:sms} reports the selected settings.

\begin{table}[t]
\centering\small
\caption{Selected DeepEP settings per schedule, topology, and phase: dispatch/combine transfer SMs, dispatch/combine epilogue SMs, and compute SM margin $m$. ``auto'' is DeepEP's automatic transfer budget (20 SMs at EP=4, 12 at EP=8); ``all'' denotes all 132 SMs.}
\label{tab:sms}
\setlength{\tabcolsep}{3pt}
\begin{tabular}{llcccc}
\toprule
& & \multicolumn{2}{c}{EP=4} & \multicolumn{2}{c}{EP=8} \\
\cmidrule(lr){3-4}\cmidrule(lr){5-6}
Schedule & & Transfer & Epil.\ / $m$ & Transfer & Epil.\ / $m$ \\
\midrule
\multirow{2}{*}{Sequential} & Prefill & 64/64 & all/all / 0 & 24/64 & all/all / 0 \\
& Decode & 8/32 & 64/64 / 0 & 2/16 & 64/64 / 0 \\
\midrule
\multirow{2}{*}{Parallel} & Prefill & 12/96 & 32/all / 0 & 24/64 & all/all / 48 \\
& Decode & 8/20 & 64/64 / 0 & 2/16 & all/all / 0 \\
\midrule
\multirow{2}{*}{Ladder} & Prefill & 64/24 & all/all / 0 & 24/24 & all/all / 32 \\
& Decode & auto/4 & 64/64 / 0 & 4/32 & all/all / 64 \\
\midrule
\multirow{2}{*}{BRAID} & Prefill & 12/64 & all/all / 16 & 24/64 & all/all / 64 \\
& Decode & 4/32 & 64/64 / 0 & 4/32 & all/all / 64 \\
\bottomrule
\end{tabular}
\end{table}

\subsection{Timing and Communication Overlap}
\label{app:inference-timing}
\paragraph{Timing.}
Each job runs two warm-up iterations per schedule, followed by 30 rounds in which every schedule runs one iteration in a randomized order. To isolate scheduling effects from differences in expert assignments and load balance, all schedules replay the same expert indices and routing weights recorded by an untimed Sequential reference run. An iteration synchronizes all GPUs, prefills one request per local prompt in one step, and generates 32 tokens. Prefill time is the first step, and decode time is the mean of the remaining 31 steps, both measured on the slowest GPU. Each job reports the median over its rounds, and we report the mean $\pm$ standard error of five job medians per workload, with speedups computed from unrounded means. Overlap measurements are medians of three Nsight Systems traces, each from a separate profiling job excluded from the latency measurements. Table~\ref{tab:timing} gives the Sequential phase timings.

\paragraph{Idealized overlap limit.}
Let $T_c$ and $T_m$ denote the computation and communication times of a sequential baseline, so that $T=T_c+T_m$. We assume that all communication can be overlapped, but only part of the computation, $T_c^{\mathrm{ov}}\leq T_c$, provides an independent overlap window. Assuming perfect overlap and no resource contention, the maximum latency reduction is $\min(T_m,T_c^{\mathrm{ov}})$, giving the ideal speedup bound
\begin{equation}
S\leq \frac{T}{T-\min(T_m,T_c^{\mathrm{ov}})}.
\label{eq:bound}
\end{equation}
For fixed $T$, the bound is larger when $T_m$ and $T_c^{\mathrm{ov}}$ are balanced, and when their minimum represents a larger fraction of $T$.

For Table~\ref{tab:timing}, define $\rho=T_m/T_c^{\mathrm{ov}}$ and $S_{\mathrm{ideal}}=T/[T-\min(T_m,T_c^{\mathrm{ov}})]$. The ratio $\rho$ identifies whether communication or the independent compute window is longer; $S_{\mathrm{ideal}}$ is an idealized overlap limit for a fixed sequential cost breakdown rather than a measured BRAID speedup.

\begin{table*}[t]
\centering\footnotesize
\caption{Sequential phase timings on rank 0, median of three Nsight Systems traces, in ms. $T$ is the total GPU span, $T_m$ is the summed duration of DeepEP's dispatch and combine kernels and their epilogues, and $T_E$ is expert computation time. We estimate the independent computation window $T_c^{\mathrm{ov}}$ from the summed durations of attention and other non-expert computation kernels. The last two columns of each phase show $\rho=T_m/T_c^{\mathrm{ov}}$ and the idealized limit $S_{\mathrm{ideal}}$ from Eq.~\eqref{eq:bound}. Traced totals are within approximately 5\% of benchmark latencies, which are measured on the slowest GPU across five separate randomized timing jobs.}
\label{tab:timing}
\setlength{\tabcolsep}{3pt}
\begin{tabular}{llrrrrcrrrrrcr}
\toprule
& & \multicolumn{6}{c}{Prefill} & \multicolumn{6}{c}{Decode step} \\
\cmidrule(lr){3-8}\cmidrule(lr){9-14}
& Workload & $T$ & $T_m$ & $T_c^{\mathrm{ov}}$ & $T_E$ & $\rho$ & $S_{\mathrm{ideal}}$ & $T$ & $T_m$ & $T_c^{\mathrm{ov}}$ & $T_E$ & $\rho$ & $S_{\mathrm{ideal}}$ \\
\midrule
EP=4 & $16\times1024$ & 91.0 & 28.7 & 20.8 & 29.6 & 1.38$\times$ & 1.30$\times$ & 9.58 & 2.16 & 2.98 & 3.78 & 0.72$\times$ & 1.29$\times$ \\
EP=4 & $64\times256$ & 87.4 & 26.9 & 19.2 & 29.2 & 1.41$\times$ & 1.28$\times$ & 10.74 & 2.12 & 3.11 & 4.86 & 0.68$\times$ & 1.25$\times$ \\
EP=4 & $64\times1024$ & 293.3 & 88.2 & 82.1 & 111.7 & 1.07$\times$ & 1.39$\times$ & 11.22 & 2.14 & 3.52 & 4.94 & 0.61$\times$ & 1.24$\times$ \\
EP=4 & $64\times4096$ & 1274.7 & 333.9 & 465.7 & 464.4 & 0.72$\times$ & 1.35$\times$ & 12.87 & 2.20 & 5.10 & 4.96 & 0.43$\times$ & 1.21$\times$ \\
EP=4 & $256\times1024$ & 1149.3 & 347.5 & 334.9 & 455.3 & 1.04$\times$ & 1.41$\times$ & 14.05 & 2.34 & 5.47 & 5.63 & 0.43$\times$ & 1.20$\times$ \\
\midrule
EP=8 & $32\times1024$ & 145.1 & 82.8 & 21.2 & 28.8 & 3.91$\times$ & 1.17$\times$ & 11.74 & 5.30 & 3.08 & 2.68 & 1.72$\times$ & 1.36$\times$ \\
EP=8 & $128\times256$ & 145.0 & 83.1 & 19.5 & 28.4 & 4.25$\times$ & 1.16$\times$ & 12.98 & 5.99 & 3.20 & 3.08 & 1.87$\times$ & 1.33$\times$ \\
EP=8 & $128\times1024$ & 481.6 & 273.6 & 82.8 & 114.6 & 3.30$\times$ & 1.21$\times$ & 13.54 & 6.00 & 3.63 & 3.14 & 1.65$\times$ & 1.37$\times$ \\
EP=8 & $128\times4096$ & 1927.6 & 978.4 & 463.0 & 468.8 & 2.11$\times$ & 1.32$\times$ & 15.16 & 6.00 & 5.22 & 3.18 & 1.15$\times$ & 1.52$\times$ \\
EP=8 & $512\times1024$ & 1844.5 & 1017.0 & 338.8 & 468.2 & 3.00$\times$ & 1.23$\times$ & 19.52 & 8.86 & 5.61 & 4.38 & 1.58$\times$ & 1.40$\times$ \\
\bottomrule
\end{tabular}
\end{table*}

For example, at EP=4 with $B=64$ and $L=1024$ in prefill, $T_m=88.2$\,ms and $T_c^{\mathrm{ov}}=82.1$\,ms yield $\rho=1.07$ and $S_{\mathrm{ideal}}=1.39$. At EP=8 with the same prompts per GPU ($B=128$), communication increases to $273.6$\,ms against an $82.8$\,ms computation window, giving $\rho=3.30$ and reducing $S_{\mathrm{ideal}}$ to $1.21$. Measured speedups remain below these limits, partly due to increased computation time: in prefill, the summed durations of BRAID's computation kernels are 17--25\% longer than Sequential's at EP=4 and 31--45\% longer at EP=8, where BRAID uses a 64-SM compute margin. These differences are consistent with resource contention and reduced compute SM availability, although differences in kernel fusion may also contribute. Thus, extensive communication overlap does not necessarily translate into a proportional improvement in end-to-end latency.

\subsection{Workload Sweeps}
\label{app:scaling}
Tables~\ref{tab:sweep} and~\ref{tab:sweep8} provide the complete latency, speedup, and overlap measurements for EP=4 and EP=8. The corresponding speedup plots appear in Figures~\ref{fig:sweeps} and~\ref{fig:sweeps8}.

\begin{table*}[t]
\centering\small
\caption{All workloads with EP=4: average of the medians of five jobs of 30 rounds $\pm$ its standard error, speedup over Sequential, and combine/dispatch overlap from Nsight Systems traces (fraction of communication-kernel time concurrent with computation on the same GPU, across GPUs).}
\label{tab:sweep}
\setlength{\tabcolsep}{3pt}
\begin{tabular}{llrr!{\tabdashvline}rrrr!{\tabdashvline}rr}
\toprule
& & \multicolumn{4}{c}{Prefill} & \multicolumn{4}{c}{Decode} \\
\cmidrule(lr){3-6}\cmidrule(lr){7-10}
Workload & Schedule & \shortstack{Latency\\(ms)} & \multicolumn{1}{r!{\tabdashvline[15pt]}}{Speedup} & \shortstack{Combine\\overlap}& \shortstack{Dispatch\\overlap} & \shortstack{Latency\\(ms/step)} & \multicolumn{1}{r!{\tabdashvline[15pt]}}{Speedup} & \shortstack{Combine\\overlap} & \shortstack{Dispatch\\overlap} \\
\midrule
\multirow{4}{*}{$16\times1024$}
& Sequential & $87.5 \pm 0.2$ & 1.000$\times$ & 0\% & 0\% & $9.36 \pm 0.01$ & 1.000$\times$ & 0\% & 0\% \\
& Parallel & $79.9 \pm 0.3$ & 1.094$\times$ & 0\% & 74.7\% & $8.60 \pm 0.01$ & 1.088$\times$ & 0\% & 86.5\% \\
& Ladder & $83.2 \pm 0.1$ & 1.052$\times$ & 81.2\% & 8.2\% & $8.41 \pm 0.01$ & 1.113$\times$ & 89.4\% & 10.0\% \\
& BRAID & $\mathbf{76.7} \pm 0.3$ & \textbf{1.141$\times$} & 86.2\% & 48.1\% & $\mathbf{7.65} \pm 0.01$ & \textbf{1.223$\times$} & 87.5\% & 77.7\% \\
\midrule
\multirow{4}{*}{$64\times256$}
& Sequential & $84.5 \pm 0.1$ & 1.000$\times$ & 0\% & 0\% & $10.64 \pm 0.01$ & 1.000$\times$ & 0\% & 0\% \\
& Parallel & $78.1 \pm 0.1$ & 1.081$\times$ & 0\% & 66.7\% & $9.86 \pm 0.01$ & 1.079$\times$ & 0\% & 85.8\% \\
& Ladder & $80.8 \pm 0.1$ & 1.045$\times$ & 81.9\% & 8.8\% & $9.54 \pm 0.01$ & 1.115$\times$ & 91.2\% & 11.5\% \\
& BRAID & $\mathbf{75.3} \pm 0.2$ & \textbf{1.122$\times$} & 89.8\% & 40.7\% & $\mathbf{8.84} \pm 0.01$ & \textbf{1.203$\times$} & 88.4\% & 81.6\% \\
\midrule
\multirow{4}{*}{$64\times1024$}
& Sequential & $291.9 \pm 0.2$ & 1.000$\times$ & 0\% & 0\% & $11.13 \pm 0.01$ & 1.000$\times$ & 0\% & 0\% \\
& Parallel & $262.5 \pm 0.5$ & 1.112$\times$ & 0\% & 93.2\% & $10.42 \pm 0.01$ & 1.068$\times$ & 0\% & 86.9\% \\
& Ladder & $267.8 \pm 0.2$ & 1.090$\times$ & 86.1\% & 11.6\% & $10.09 \pm 0.01$ & 1.103$\times$ & 92.8\% & 11.1\% \\
& BRAID & $\mathbf{249.2} \pm 0.4$ & \textbf{1.171$\times$} & 96.3\% & 92.8\% & $\mathbf{9.41} \pm 0.01$ & \textbf{1.182$\times$} & 89.2\% & 82.7\% \\
\midrule
\multirow{4}{*}{$64\times4096$}
& Sequential & $1279 \pm 2$ & 1.000$\times$ & 0\% & 0\% & $12.92 \pm 0.01$ & 1.000$\times$ & 0\% & 0\% \\
& Parallel & $1158 \pm 1$ & 1.105$\times$ & 0\% & 99.7\% & $12.18 \pm 0.01$ & 1.061$\times$ & 0\% & 85.5\% \\
& Ladder & $1186 \pm 1$ & 1.078$\times$ & 86.7\% & 13.2\% & $11.83 \pm 0.01$ & 1.092$\times$ & 95.4\% & 11.1\% \\
& BRAID & $\mathbf{1101} \pm 2$ & \textbf{1.161$\times$} & 98.0\% & 99.6\% & $\mathbf{11.14} \pm 0.01$ & \textbf{1.160$\times$} & 90.4\% & 88.4\% \\
\midrule
\multirow{4}{*}{$256\times1024$}
& Sequential & $1154 \pm 2$ & 1.000$\times$ & 0\% & 0\% & $14.01 \pm 0.01$ & 1.000$\times$ & 0\% & 0\% \\
& Parallel & $1052 \pm 3$ & 1.096$\times$ & 0\% & 93.0\% & $13.23 \pm 0.01$ & 1.059$\times$ & 0\% & 89.1\% \\
& Ladder & $1061 \pm 2$ & 1.088$\times$ & 85.6\% & 13.1\% & $12.83 \pm 0.01$ & 1.092$\times$ & 95.9\% & 10.8\% \\
& BRAID & $\mathbf{986} \pm 2$ & \textbf{1.170$\times$} & 98.0\% & 94.5\% & $\mathbf{12.33} \pm 0.01$ & \textbf{1.136$\times$} & 89.6\% & 89.6\% \\
\bottomrule
\end{tabular}
\end{table*}

\begin{table*}[t]
\centering\small
\caption{All workloads with EP=8 on two nodes: average of the medians of five jobs of 30 rounds $\pm$ its standard error, speedup over Sequential, and combine/dispatch overlap from Nsight Systems traces (fraction of communication-kernel time concurrent with computation on the same GPU, across GPUs).}
\label{tab:sweep8}
\setlength{\tabcolsep}{3pt}
\begin{tabular}{llrr!{\tabdashvline}rrrr!{\tabdashvline}rr}
\toprule
& & \multicolumn{4}{c}{Prefill} & \multicolumn{4}{c}{Decode} \\
\cmidrule(lr){3-6}\cmidrule(lr){7-10}
Workload & Schedule & \shortstack{Latency\\(ms)} & \multicolumn{1}{r!{\tabdashvline[15pt]}}{Speedup} & \shortstack{Combine\\overlap} & \shortstack{Dispatch\\overlap} & \shortstack{Latency\\(ms/step)} & \multicolumn{1}{r!{\tabdashvline[15pt]}}{Speedup} & \shortstack{Combine\\overlap} & \shortstack{Dispatch\\overlap} \\
\midrule
\multirow{4}{*}{$32\times1024$}
& Sequential & $141.1 \pm 0.1$ & 1.000$\times$ & 0\% & 0\% & $11.69 \pm 0.02$ & 1.000$\times$ & 0\% & 0\% \\
& Parallel & $123.4 \pm 0.1$ & 1.143$\times$ & 0\% & 48.5\% & $9.67 \pm 0.02$ & 1.208$\times$ & 0\% & 83.5\% \\
& Ladder & $126.7 \pm 0.1$ & 1.113$\times$ & 75.2\% & 2.5\% & $10.45 \pm 0.02$ & 1.118$\times$ & 59.9\% & 4.7\% \\
& BRAID & $\mathbf{122.7} \pm 0.1$ & \textbf{1.150$\times$} & 88.6\% & 19.1\% & $\mathbf{9.59} \pm 0.02$ & \textbf{1.219$\times$} & 82.6\% & 4.4\% \\
\midrule
\multirow{4}{*}{$128\times256$}
& Sequential & $137.9 \pm 0.4$ & 1.000$\times$ & 0\% & 0\% & $12.83 \pm 0.02$ & 1.000$\times$ & 0\% & 0\% \\
& Parallel & $122.2 \pm 0.4$ & 1.128$\times$ & 0\% & 43.1\% & $10.68 \pm 0.02$ & 1.201$\times$ & 0\% & 76.6\% \\
& Ladder & $124.6 \pm 0.4$ & 1.107$\times$ & 70.3\% & 2.6\% & $11.20 \pm 0.03$ & 1.146$\times$ & 55.7\% & 4.4\% \\
& BRAID & $\mathbf{121.3} \pm 0.4$ & \textbf{1.136$\times$} & 87.9\% & 15.9\% & $\mathbf{10.24} \pm 0.04$ & \textbf{1.253$\times$} & 79.3\% & 4.5\% \\
\midrule
\multirow{4}{*}{$128\times1024$}
& Sequential & $479 \pm 3$ & 1.000$\times$ & 0\% & 0\% & $13.45 \pm 0.03$ & 1.000$\times$ & 0\% & 0\% \\
& Parallel & $408 \pm 3$ & 1.175$\times$ & 0\% & 52.9\% & $11.01 \pm 0.03$ & 1.222$\times$ & 0\% & 90.1\% \\
& Ladder & $423 \pm 3$ & 1.133$\times$ & 90.6\% & 2.6\% & $11.55 \pm 0.04$ & 1.165$\times$ & 65.2\% & 4.3\% \\
& BRAID & $\mathbf{407} \pm 2$ & \textbf{1.178$\times$} & 95.3\% & 31.3\% & $\mathbf{10.53} \pm 0.03$ & \textbf{1.278$\times$} & 88.8\% & 5.8\% \\
\midrule
\multirow{4}{*}{$128\times4096$}
& Sequential & $1940 \pm 10$ & 1.000$\times$ & 0\% & 0\% & $15.05 \pm 0.06$ & 1.000$\times$ & 0\% & 0\% \\
& Parallel & $1530 \pm 8$ & 1.271$\times$ & 0\% & 81.9\% & $12.52 \pm 0.03$ & 1.202$\times$ & 0\% & 95.1\% \\
& Ladder & $1740 \pm 10$ & 1.120$\times$ & 98.1\% & 2.8\% & $12.27 \pm 0.04$ & 1.226$\times$ & 93.5\% & 4.0\% \\
& BRAID & $\mathbf{1510} \pm 10$ & \textbf{1.287$\times$} & 97.1\% & 69.1\% & $\mathbf{10.98} \pm 0.08$ & \textbf{1.370$\times$} & 91.9\% & 46.8\% \\
\midrule
\multirow{4}{*}{$512\times1024$}
& Sequential & $1833 \pm 9$ & 1.000$\times$ & 0\% & 0\% & $19.58 \pm 0.09$ & 1.000$\times$ & 0\% & 0\% \\
& Parallel & $1541 \pm 6$ & 1.190$\times$ & 0\% & 54.0\% & $16.20 \pm 0.08$ & 1.209$\times$ & 0\% & 86.8\% \\
& Ladder & $1615 \pm 7$ & 1.135$\times$ & 88.4\% & 2.7\% & $15.75 \pm 0.09$ & 1.244$\times$ & 84.6\% & 3.2\% \\
& BRAID & $\mathbf{1528} \pm 7$ & \textbf{1.200$\times$} & 97.2\% & 36.1\% & $\mathbf{14.35} \pm 0.09$ & \textbf{1.365$\times$} & 93.5\% & 22.7\% \\
\bottomrule
\end{tabular}
\end{table*}

\begin{figure*}[t]
    \centering
    \includegraphics[
        width=\textwidth
    ]{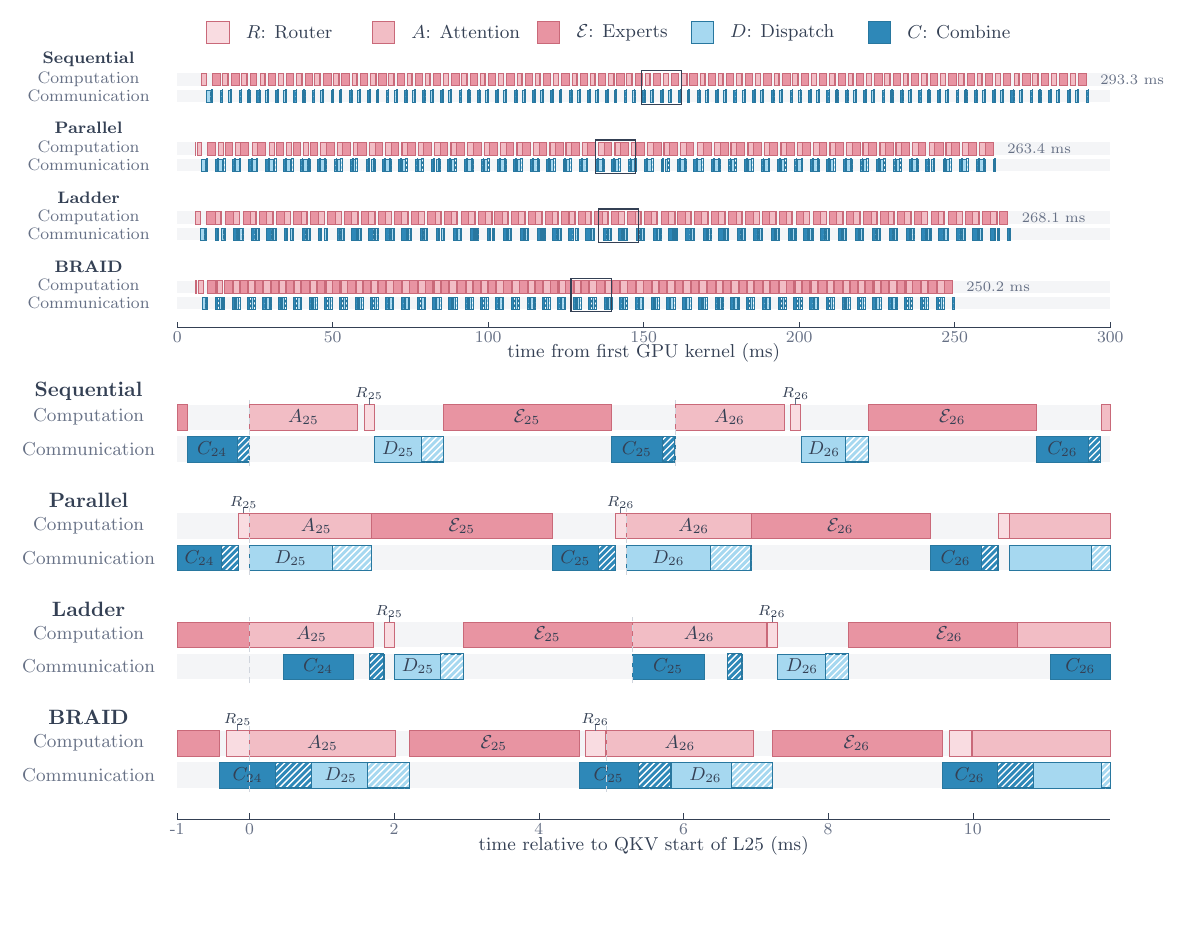}
\caption{GPU execution timelines for Qwen3-30B-A3B-Base in BF16 with vLLM and DeepEP on four H100 GPUs (EP${}={}$4), using a global batch of 64 prompts of length 1,024. The upper timelines show the complete 48-layer prefill, while the lower timelines zoom in on layers 25 and 26 together with the preceding layer-24 combine. Computation blocks span from the first to the last GPU kernel of each logical phase, while each DeepEP dispatch or combine transfer and its GPU epilogue are shown separately. Epilogues retain the color of their corresponding communication phase and are marked with diagonal white stripes. Small unlabeled intervals contain residual or normalization work, or execution gaps. These timelines come from a single Nsight-instrumented iteration on rank 0; Table~\ref{tab:perf} reports the randomized latency measurements.}
    \label{fig:vllm-execution-traces}
\end{figure*}

\end{document}